\documentclass[onecolumn,12pt]{article}
\pdfoutput=1
\usepackage{jheppub}

\usepackage[font={small}]{caption}
\usepackage{subfigure}
\usepackage{graphicx}
\usepackage{dcolumn}
\usepackage{float}
\usepackage[normalem]{ulem}

\usepackage{mathrsfs} 

\usepackage{multicol}
\usepackage{cancel}
\usepackage{mathtools}
\usepackage{hyperref}

\usepackage{subfigure}
\usepackage{xcolor}
\def\({\left(} \def\){\right)}
\def\[{\left[} \def\]{\right]}

\newcommand{\eg}{{\it e.g.,}\ }
\newcommand{\ie}{{\it i.e.,}\ }

\newcommand{\R}{{\mathbb R}}
\newcommand{\bea}{\begin{eqnarray}}
\newcommand{\eea}{\end{eqnarray}}

\renewcommand{\eqref}[1]{(\ref{#1})}

\begin{document}


%
\title{Holographic Complexity and de Sitter Space}

\author{Shira Chapman$^{1}$, Dami\'an A. Galante$^2$, and Eric David Kramer$^3$}

\affiliation{$^1$ Department of Physics, Ben-Gurion University of the Negev, Beer Sheva 84105, Israel}
\affiliation{$^2$ Department of Mathematics, King’s College London, the Strand, London WC2R 2LS, UK}
\affiliation{$^3$ Racah Institute of Physics, Hebrew University of Jerusalem, Jerusalem 91904, Israel}

\emailAdd{schapman@bgu.ac.il}\emailAdd{damian.galante@kcl.ac.uk}
\emailAdd{ericdavidkramer@gmail.com}

%
%
%
%
%
%
%
%
%

\abstract{
\sloppy We compute the length of spacelike geodesics anchored at opposite sides of certain double-sided flow geometries in two dimensions. These geometries are asymptotically anti-de Sitter but they admit either a de Sitter or a black hole event horizon in the interior. While in the geometries with black hole horizons, the geodesic length always exhibit linear growth at late times, in the flow geometries with de Sitter horizons, geodesics with finite length only exist for short times of the order of the inverse temperature and they do not exhibit linear growth. We comment on the implications of these results towards understanding the holographic proposal for quantum complexity and the holographic nature of the de Sitter horizon.}

%
%

\maketitle

\setcounter{tocdepth}{2}

\section{Introduction}
The objective of this paper is to study cosmological event horizons from a modern holographic perspective. Cosmological horizons surround observers in universes with a positive cosmological constant, like the one our own universe asymptotes to.  Despite their obvious relevance, they have been much less explored compared to black hole horizons. Cosmological horizons behave differently from black hole horizons, see \eg \cite{Anninos:2012qw,Spradlin:2001pw}, even though, for instance, their entropy is also proportional to the area of the event horizon \cite{Gibbons:1977mu}. 

The AdS/CFT correspondence \cite{Aharony:1999ti} provides a fruitful framework for studying black hole horizons in negatively curved spaces using conformal field theory (CFT) observables located at the boundary. In this arena, a special role is played by tools from the world of quantum information, see \eg \cite{RevModPhys.88.015002,rangamani2017holographic,Susskind:2018pmk}. It would be surprising if similar tools did not play a central role in understanding the cosmological horizon as well. 
However, the fact that there is no timelike boundary in de Sitter spacetime (dS) is an obstacle in translating ideas from gravitational holography into the cosmological case. 

\sloppy
Past attempts to study de Sitter holographically include the  dS/CFT correspondence \cite{Strominger:2001pn,Witten:2001kn,Strominger:2001gp,Maldacena:2002vr,Anninos:2011ui,Anninos:2017eib},
thinking of the dS horizon as a holographic screen, \eg  \cite{Banks:2005bm,Banks:2006rx,banks2012holographic} and static patch holography which associates a quantum mechanical model with the observer's worldline \cite{Anninos:2011af,Leuven:2018ejp}. 
Recently, a new set of ideas appeared including the use of $T\bar{T}$ deformations 
\cite{Gorbenko:2018oov,Lewkowycz:2019xse,Shyam:2021ciy} and a cosmological bootstrap program \cite{Arkani-Hamed:2018kmz,Hogervorst:2021uvp,DiPietro:2021sjt}.
It is fair to say, though, that still there is no single microscopic quantum model to describe the cosmological horizon.

Interest in a holographic description of dS is further motivated by the fact that it has been alarmingly difficult to find stable de Sitter-like vacua in String Theory. The few successful attempts such as \cite{Kachru_2003} are still a matter of controversy. Some authors have gone so far as to conjecture that there do not exist stable de Sitter-like vacua in String Theory \cite{obied2018sitter}. While it is indeed important to continue the search for stable vacua, further study of holography, especially the question of whether a holographic description of dS is even possible, is an alternative way of exploring quantum gravity in dS.

Here, we continue the effort started in \cite{Anninos:2017hhn,Anninos:2018svg}  to probe the cosmological horizon using the standard tools of the AdS/CFT correspondence. The main idea behind this program is to embed part of a dS universe inside AdS and, in doing so, providing a boundary to study the cosmological horizon. Embedding dS$_{d+1}$ inside AdS$_{d+1}$ in $d>1$ was first attempted in \cite{Freivogel:2005qh,Lowe:2010np}, where it was observed that in order to satisfy the null energy condition, it was necessary to hide the dS patch inside a black hole horizon in AdS. However, a new set of geometries was proposed in \cite{Anninos:2017hhn,Anninos:2018svg}, where the cosmological event horizon is in causal contact with the AdS boundary. These are solutions to certain dilaton-gravity theories in two dimensions that, when uplifted to higher dimensions, do satisfy the null energy conditions \cite{Anninos:2020cwo}. They appear in the context of the recently studied near AdS$_2$ geometries \cite{Almheiri:2014cka,Jensen:2016pah,Maldacena:2016upp,Engelsoy:2016xyb}, where there is a large, slowly varying dilaton playing the role of the size of the compact dimensions. 
These geometries share similar features to the low energy regime of the SYK model \cite{Maldacena:2016hyu,Sarosi:2017ykf}. From this point of view, they can be seen as an RG flow from a UV near-conformal point towards a dS infrared point and we therefore refer to them as flow geometries. One could imagine building the dual to the flow geometries from relevant deformations of SYK-like models \cite{Anninos:2020cwo}.

From a macroscopic perspective, the flow geometries allow us to compute different types of observables in the hope of characterising the cosmological event horizon. These observables turn out to differ significantly from their counterparts for black hole horizons. Examples of these include: the frequencies of the dissipative quasinormal modes, which have a small real part which indicates that the geometry in the deconfined phase is less efficient at thermalizing  \cite{Anninos:2017hhn}; the out-of-time-ordered four point function, that oscillates in time rather than obeying the acclaimed exponential growth of chaotic systems  \cite{Anninos:2018svg}; and, 
positive energy shockwaves, which open the wormhole rather than closing it \cite{Gao:2000ga}. All these pose challenges in the microscopic interpretation of the flow  geometries. 

Furthermore, in three dimensions, corrections to the cosmological horizon entropy were recently computed, finding again notable differences with respect to analogous corrections for black holes \cite{Anninos:2020hfj,Anninos:2021ihe}.

In this paper, we concentrate on a different macroscopic observable: the volume of an extremal spacelike codimension-one slice connecting opposite sides of an eternal double-sided geometry. 
In two dimensions, this is simply the length of a geodesic. This observable has recently gained a lot of attention due to its ability to probe the horizon interior and also due to its connection to the notion of quantum computational  complexity.

Quantum computational complexity is a notion from quantum information which estimates the difficulty of constructing a quantum state from simple elementary operations \cite{watrous2009quantum,aaronson2016complexity}. Complexity has some striking features which distinguish it from other measures of quantum correlations. Specifically, in chaotic systems, the complexity grows linearly following a quantum quench for a long time (exponential in the entropy of the system) and then saturates. It also reacts to perturbations in a characteristic way which encodes chaos and scrambling. All these behaviours have been reproduced using the maximal volume slices in AdS black holes.  
see \eg \cite{Susskind:2014rva,Stanford:2014jda,Susskind:2014moa,Carmi:2017jqz,Chapman:2018dem,Chapman:2018lsv,Iliesiu:2021ari}.

These similarities led to conjecture that complexity of a quantum state is a plausible holographic dual to the extremal volume anchored at the boundary times where the state is defined
\begin{equation}\label{CV}
\mathcal{C}_V = \max \frac{V}{G_N \ell}
\end{equation}
where $\ell$ is a certain length scale associated with the geometry, usually selected to be the AdS radius of curvature. In two-dimensional dilaton gravity it was suggested in \cite{Brown:2018bms} that equation \eqref{CV} should include an additional factor $\Phi_0$ which is the constant part of the full dilaton field. 
Alternatively, propositions were made which relate the complexity to the action of the WdW patch \cite{Brown:2015lvg,Brown:2015bva} and to its spacetime volume \cite{Couch:2016exn}.
In cases with a horizon, all these quantities probe the behind horizon region.   
Here we explore the complexity=volume (CV) conjecture in flow geometries.

We find that the geodesics in the flow geometries are strongly affected by the interior dS$_2$ region and 
 differ significantly from geodesics in AdS$_2$ black holes.  
It is well known that  spacelike geodesics in dS starting at a point will reach its antipodal point and that not all points on the worldline of an observer are connected by a spacelike geodesic with points on the ``antipodal worldline''. As a consequence, we will see that in most of our flow geometries not all boundary times will be connected by  spacelike geodesics and only a finite short range of times of the order of the inverse temperature will have a complexity=volume observable associated to it.

Another important difference is that, even when they exist, the length of geodesics does not behave as in the case of black holes. 
In the usual limit where the boundary lies very far from the horizon, the length of geodesics in the flow geometries with dS horizons behave as\footnote{Slightly different behaviours can be obtained with different types of flow geometries with dS interiors as $c_\gamma$ becomes imaginary, but since the solutions are valid only for a short range of times, none of them produces linear growth at late times. See section \ref{gamma_cen_sec}.}  
\begin{equation}\label{intro:AdS}
\mathcal{C}_V(t) \sim S_0 \log \, \cos (c_\gamma \pi  T t) + \text{const}\,, 
\end{equation}
where $c_\gamma$ depends on the specific flow geometry we consider and  $S_0$ is the entropy associated with the constant part of the dilaton. In most cases, the volume complexity decreases at early times reaching a minimum. This behaviour is valid only for times of the order the of the inverse temperature. At that point, there is a last geodesic that becomes (almost everywhere) null and reaches past/future infinity. At later times, finite-length geodesics cease to exist. A similar phenomenon was observed for pure dS geometries in \cite{Susskind:2021esx}. 
This contrasts with the known result for the AdS$_2$ black hole, 
\begin{equation}\label{intro:centaur}
\mathcal{C}_V(t) \sim S_0   \log\, \cosh (\pi T t) + \text{const}, 
\end{equation}
where the volume complexity grows linearly at late times  $tT\gg1$. 

One might suspect that the different behaviours come from having glued together two geometries. In order to rule this out, we consider flow geometries that interpolate between an AdS black hole in the interior and an AdS with different curvature radius close to the boundary. In this case, we do recover exactly the same linear growth at late times as in the black hole case.

The rest of the paper is organised as follows: in section \ref{2d}, we present the dilaton-gravity theories under consideration; in section \ref{geodesics_2d}, we build the formalism to compute the length of the geodesics for an arbitrary geometry; section \ref{sec_max} discusses the known examples of some maximally symmetric spacetimes through this formalism; in section \ref{sec_flow}, we compute the lengths of geodesics in different flow geometries that interpolate between an AdS boundary and  different (A)dS interiors; we end up with a discussion of the different results in section \ref{discussion}. Some details of the flow geometries have been relegated to two appendices.

\section{2d dilaton-gravity theories} \label{2d}

We will study general dilaton-gravity theories in two dimensions. The Lorentzian action is given by, 
\begin{equation}
\begin{split}
S = &\frac{\Phi_0}{16 \pi G_N} \left(\int d^2x \sqrt{-g} R +
 2 \int du \sqrt{-h} K \right)
\\
&+  \frac{1}{16 \pi G_N} \int d^2x \sqrt{-g}  \left( \phi  R + \ell^{-2} U(\phi) \right) + \frac{1}{8\pi G_N} \int du \sqrt{-h} \phi_b (K-1/\ell) \,. 
\end{split}
\end{equation}
The first term is topological 
and proportional to the Euler characteristic of the manifold.\footnote{Here we only focus on solutions which have a trivial topology, but see, \eg \cite{Iliesiu:2021ari} for the influence of different topologies on the complexity.} $K$ and $h$ are the extrinsic curvature and the induced metric on the boundary, respectively, $\ell$ is the curvature radius of the manifold,  and $\phi_b$ is the value of the dilaton at the boundary.  
 Finally, we require the full dilaton to be positive $\Phi = \Phi_0+\phi>0$ and we work in the limit where $\Phi_0 \gg \phi$. 

The equations of motion for the dilaton and the metric read
\begin{equation}
\begin{split}\label{JTJTEOM}
R & =  - \frac{U'(\phi) }{\ell^2} \,, \\
0 & =  \nabla_a \nabla_b \phi - g_{ab} \nabla^2 \phi + \frac{g_{ab}}{2\ell^2} U(\phi)  \,.
\end{split}
\end{equation}
Examples of such theories include the JT gravity theory where the dilaton potential is set to $U(\phi) = 2\phi$. For any sufficiently smooth dilaton potential, the equations of motion \eqref{JTJTEOM} admit solutions given by
\begin{equation}
ds^2 = -f(r) dt^2 + \frac{dr^2}{f(r)} \,\,\,\,\,\,\, , \,\,\,\,\,\,\, \phi = r /\ell \,,  \label{2dmetric}
\end{equation}
where the blackening factor $f(r)$ is
\begin{equation}\label{blackfactor}
f(r) = \int_{r_h/\ell}^{r/\ell} U(\phi) d\phi \,,
\end{equation}
and $r_h$ is the position of the event horizon where the blackening factor vanishes. It is straightforward to check that this solution satisfies the equations of motion for any potential. The thermodynamics of these dilaton gravity theories was studied in \cite{Anninos:2017hhn} (see also Appendix \ref{app_thermo}), where it was demonstrated that the temperature and entropy are given by 
\begin{equation}\label{thermoquants}
T = \frac{U(\phi(r_h))}{4\pi \ell}~, \quad\quad S = \frac{\Phi_0 + \phi(r_h)}{4 G_N } ~.
\end{equation}

The following derivations will not use the specific form of the blackening factor $f(r)$ from equation \eqref{blackfactor}. Instead, we will use only the fact that our Lorentzian geometry takes the form \eqref{2dmetric} where 
$t \in \R$ and $f(r)$, for now, is a generic continuous function of the $r$-coordinate with AdS asymptotics, \ie $f(r) \to r^2$ when $r \to \infty$.\footnote{In section \ref{sec_ds}, we will relax this assumption when considering purely dS spacetime.} We further assume that $f(r)$ has a single root within the physical range of the coordinate $r$ at some $r=r_h$ indicating the position of the horizon, that it is positive outside the horizon, and that the geometry can be maximally extended into a two-sided geometry with two boundaries.

\begin{figure}[h]
	\centering
	\includegraphics[scale=1]{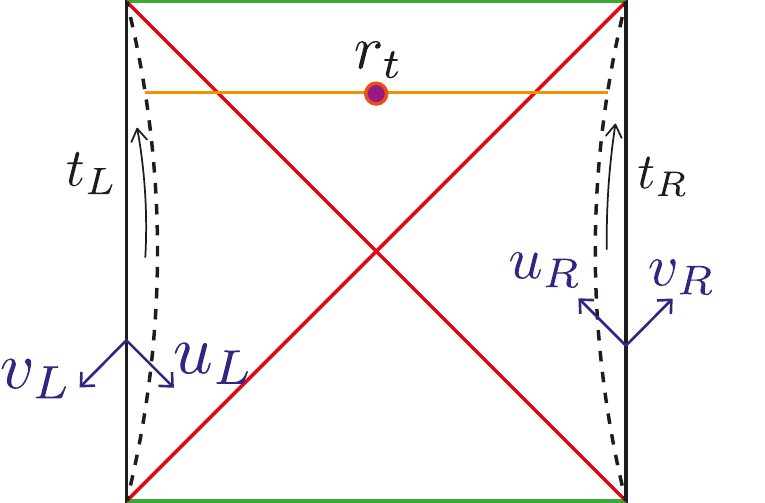}
	\caption{Penrose diagram for AdS$_2$. The boundary $r=R_b$, corresponding to some fixed value of the dilaton $\phi=\phi_b$, is indicated by a dashed black line. The times $t_L=t_R$ run upwards along both boundaries. The axis of changing $u_{L/R},v_{L/R}$ are indicated in the figure. We have also illustrated a geodesic with turning point $r_t$ (see below).}
	\label{fig:penroseperlim}
\end{figure}

\fussy
In order to extend the geometry, it is useful to define a tortoise coordinate,\footnote{Some care has to be taken when evaluating the tortoise coordinate across the horizon. Under our assumptions, the blackening factor takes the form $f(r)=(r-r_h) F(r)$ where $F(r)$ has no other roots within the physical range of $r$. The inverse of $f(r)$ can be decomposed as 
\begin{equation}\label{decomblack}
\frac{1}{f(r)} = \frac{1}{F(r_h)(r-r_h)}+\frac{F(r_h)-F(r)}{F(r)F(r_h)(r-r_h)}
\end{equation}
which integrates to
\begin{equation}
r^*(r) = \log\frac{|r-r_h|}{F(r_h)}+G(r)
\end{equation}
where $G(r)$ comes from integrating the second term on the right hand side of \eqref{decomblack} and is completely regular at $r=r_h$.\label{footy4}}
\begin{equation}\label{tortoisy}
r^*(r) = \int^r \frac{d\tilde r}{f(\tilde r)} \,.
\end{equation}
Without loss of generality, we can choose the integration constant such that $r^*(r\rightarrow\infty)=0$, \ie the tortoise coordinate vanishes at the AdS boundary. 
We next define lightcone coordinates,
\begin{eqnarray}\label{lightcoords}
v_R  =  t_R + r^* \,, \qquad 
u_R  =  t_R - r^* \,.
\end{eqnarray}
The $R$ subscript indicates that these coordinates cover the right-side of the two-sided Penrose diagram. In these coordinates the metric (\ref{2dmetric}) becomes
\begin{equation}
ds^2 = -f(r) dv_R^2 + 2 dv_R dr = -f(r) du_R^2 - 2 du_R dr \,.
\end{equation}
Similarly, we can define a set of left coordinates, $v_L = -t_L + r^*$, $u_L = -t_L - r^*$, that cover the left part of the Penrose diagram. Note that with this choice of coordinates, the time variable runs upwards along both boundaries. The different lightcone coordinates are depicted in Fig. \ref{fig:penroseperlim}.

\subsection{Penrose diagram coordinates}
In order to draw a Penrose diagram it is convenient to define new coordinates $x^+, x^-$ that are finite across the horizon. In the $x^+,x^->0$ quadrant, these are defined by
\begin{equation}
x^+ =  e^{v_R/\ell} \,\,\,\, \,, \,\,\,\, x^- =  e^{-u_R/\ell } \,.
\end{equation}
Similar formulas apply in the other quadrants with some overall sign modifications. Note that $x^+ x^- = e^{2r^*/\ell}$, so constant $x^+ x^-$ correspond to constant-$r$ slices and similarly, constant $x^+/x^- =e^{2t_R/\ell}$ corresponds to constant-$t_R$ slices. The $x^\pm$ coordinates still run from $-\infty$ to $\infty$, so to compactify them into the Penrose diagram we define $U_R,V_R$ coordinates such that
\begin{equation}
x^+ = \tan U_R \,\,\,\, , \,\,\,\, x^- = \tan V_R \,.
\end{equation}
We will generally use coordinates where the asymptotic boundary of AdS is at $x^+ x^- = 1$, so that each boundary is a vertical line in the Penrose diagram.


\section{Geodesics in 2d spacetimes} \label{geodesics_2d}

The aim of this paper is to study the {\textit{volume}}\footnote{The volume usually refers to the size of a codimension-one surface. In this paper we are interested in two dimensional geometries, so the volume is equivalent to the length of the geodesics. We will keep using the term volume throughout the rest of the text to be consistent with the complexity literature in higher dimensions.} of spacelike geodesics that are anchored at fixed times on the two boundaries. It is possible to develop a formalism to find these geodesics and compute their volume for generic $f(r)$, see \eg \cite{Carmi:2017jqz,Chapman:2018lsv}. We explain how to do this in the current section. In the following sections, we will use this formalism to study the geodesics in specific examples including the flow geometries.

\subsection{Geodesics for general $f(r)$ geometries}\label{genfrproc}
To find geodesics in geometries described in terms of a blackening factor $f(r)$, consider the volume of these geodesics
\begin{equation}
V = \int ds \sqrt{-f \dot{v}_R^2 + 2 \dot{v}_R \dot{r}} = \int ds \sqrt{-f \dot{u}_R^2 - 2 \dot{u}_R \dot{r}} \,,
\label{volume2}
\end{equation}
where $v_R, u_R$ and $r$ are parametrized in terms of a parameter $s$ and the dot indicates derivative with respect to this parameter. It is always possible to choose a parametrization where
\begin{equation}
-f \dot{v}_R^2 + 2 \dot{v}_R \dot{r} = - f \dot{u}_R - 2 \dot{u}_R \dot{r} =1 \,,
\end{equation}
so that
\begin{eqnarray}
\dot{r} = \frac{1+ f \dot{v}_R^2}{2 \dot{v}_R} = - \frac{1+ f \dot{u}_R^2}{2 \dot{u}_R} \,.
\end{eqnarray}
The volume does not depend explicitly on the $v_R$ (or $u_R$) coordinate, so there is a conserved quantity 
\begin{equation}
P_v = \frac{\delta V}{\delta \dot{v}_R} = -f \dot{v}_R + \dot{r} = \frac{1- f \dot{v}_R^2}{2 \dot{v}_R} =-f \dot{u}_R - \dot{r} = \frac{1- f \dot{u}_R^2}{2 \dot{u}_R}=\frac{\delta V}{\delta \dot{u}_R}= P_u \equiv P \,.
\label{Peq}
\end{equation}
From here, we can solve for $\dot{v}_R$ and $\dot{u}_R$  as a function of $P$ and $r$ and obtain
\begin{equation}
\dot{v}_{R\pm} = \frac{-P \pm \sqrt{f + P^2}}{f} = \dot{u}_{R\mp} \,,
\label{dotv}
\end{equation}
which in turn implies using \eqref{Peq} that
\begin{equation}
\dot{r}_\pm = \pm \sqrt{f + P^2} \,. 
\label{dotr}
\end{equation}
Here, we see that we can interpret the subscripts $\pm$ labelling the the different solutions as an indication of whether the radial coordinate $r$ is increasing or decreasing with increasing $s$ along the geodesic.  

At points where \eqref{dotr} changes sign the geodesic will turn around, \ie if before this point the geodesic was moving away from the boundary into the interior of the geometry, after this point it will go back towards the boundary (in the second side of the double sided geometry). We denote the point where $\dot{r}_\pm = 0$ by $r_t$ and refer to it as the turning point, see figure \ref{fig:penroseperlim}. It can be obtained by solving the equation
\begin{equation}\label{turnP}
f(r_t) + P^2 = 0 \,.
\end{equation}
Except for specific degenerate cases, all the geometries considered in this paper will admit a single turning point, \ie equation \eqref{turnP} will have a single solution within the physical range of the coordinate $r$. The position of this turning point will depend on the form of $f(r)$ so we will discuss it later for each of the examples separately. In geometries with shockwaves \cite{Stanford:2014jda,Chapman:2018lsv} multiple turning points can occur, but we will not be dealing with such cases here.

Next, let us write down expressions for the volume of the extremal slices. The isometries of our geometries imply that 
 the volume is invariant under the following change of the boundary times $t_R\rightarrow t_R+\Delta t$ and $t_L \rightarrow t_L-\Delta t$ and hence only depends on the combination $t_L+t_R$. For simplicity, we will assume a symmetric configuration of the boundary times $t_L=t_R=t/2$, but our result will be valid also for non-symmetric  configurations. With the symmetric configuration,   
the total volume will be twice the volume on each side of the geometry. The latter is obtained by integrating 
\eqref{volume2} from the turning point towards the boundary along increasing $r$. We will assume that the boundary is fixed at some $r|_{bdy} = R_b$ where the dilaton takes a constant value $\phi=\phi_b=R_b/\ell$, see eq.~\eqref{2dmetric}. Then,
\begin{equation}
V[P] = \int ds = 2 \int_{r_t}^{R_b} \frac{dr}{\dot{r}_+} = 2 \int_{r_t}^{R_b} \frac{dr}{\sqrt{f(r) + P^2}} \,.
\label{volume}
\end{equation}
This integral is well-behaved for any finite $R_b$. In particular, the integrand is smooth at the horizon.

Finally, we would like to relate the volume to the boundary times at which the geodesic is anchored. 
To find an expression for the boundary times in terms of the momentum $P$, we integrate eqs.~\eqref{dotv}-\eqref{dotr} according to
\begin{eqnarray}
v_R(R_b) - v_R(r_t) & = & \int_{r_t}^{R_b} dr   \frac{\dot{v}_{R+}}{ \dot{r}_+} =  \int_{r_t}^{R_b}  \tau(P,r) dr  \,, \label{v_eq} \\
u_L (r_t) - u_L (R_b) & = &  \int_{R_b}^{r_t} dr   \frac{\dot{u}_{L-}}{ \dot{r}_-}  =  \int_{r_t}^{R_b} \tau(P,r) dr   \,,
\label{bdytime}
\end{eqnarray}
where 
\begin{equation}
\tau (P, r) \equiv  \frac{\sqrt{f(r) + P^2} - P}{f(r) \sqrt{f(r) + P^2}}  \,.
\end{equation}
In the above expressions, we have assumed that $P>0$. The geodesic moves from the left boundary to the right boundary, crossing behind the future horizon. Note that  the integrand $\tau(P,r)$ does not diverge around $r=r_h$.\footnote{We always parametrize our geodesics starting at the left boundary and ending on the right boundary. In this case $P>0$ corresponds to a geodesic crossing behind the future horizon which can be treated in terms of the $v_R$ and $u_L$ coordinates. The alternative case $P<0$ of geodesics passing behind the past horizon  should be treated using the $u_R$ and $v_L$ coordinates. In the latter case, the time integral in eq.~\eqref{time_int}  will be replaced with
\begin{equation}
\frac{t}{2} =  -r^*_t + r^* (R_b)  - \int_{r_t}^{R_b} \tau (-P,r) dr \,.
\end{equation}
Note that here too, the integrand does not diverge at the horizon. Most of the expressions we present in the following sections will be valid for both $P>0$ and $P<0$.} 
Above, we have chosen the same boundary cutoff $R_b$ for the right and left boundaries.

Using the definition of the coordinates and summing up  equations \eqref{v_eq}-\eqref{bdytime}, we end up with
\begin{equation}
\frac{t}{2} =  r^*_t - r^* (R_b)  + \int_{r_t}^{R_b} \tau (P,r) dr \,,
\label{time_int}
\end{equation}
where we have defined the total boundary time\footnote{Here $t_L$ and $t_R$ indicate the value of the time coordinate along the boundary curve $r=R_b$.}
\begin{equation}
t\equiv t_L+t_R\,.
\end{equation}
The relations \eqref{volume} and \eqref{time_int} are parametric  equations for the volume and time in terms  of the momentum $P$. 
Alternatively, inverting \eqref{time_int} gives $P(t)$, from which we can obtain $V(t)$.

Despite the somewhat complicated integrals which we will have to perform separately for each $f(r)$, it turns out that the rate of change of the volume has a very simple expression in terms of the momentum (cf. \cite{Carmi:2017jqz,Chapman:2018lsv}):
\begin{equation}\label{dVdtgen}
\frac{dV}{d(t_R + t_L)} = P  \,.
\end{equation}

\subsection{Geodesics are always maximal}
By the upper semi-continuity of arc-length \cite{Wald:1984rg}, spacelike geodesics, being extremal, will always have maximal volume, while  timelike geodesics will always have maximal proper time. In our case, we can verify explicitly that the volume of the geodesics corresponds to a maximum. Consider the volume \eqref{volume2} in a parametrization set by the radial coordinate $s=r$. The the second variation of the volume functional $V[v(r)]$ with respect to the path $v(r)$ reads
\begin{align}
	\delta^{(2)}V = -\int\!dr\,\frac{1}{\mathcal{V}^{3}}\,\delta v'(r)^2\;,
\end{align} 
where $\mathcal{V} =\sqrt{-f(r)v'(r)^2+ 2v'(r)}$ is the (positive) volume element and $\delta v'(r) = \frac{d}{dr} \delta v(r)$. (There is no term proportional to $\delta v(r)$ because the integrand only depends on $v'(r)$.) Since the expression for $\delta^{(2)}V$ is manifestly negative, we conclude that the volume of geodesic corresponds to a maximum. We will see later situations where multiple geodesics correspond to some fixed boundary times and in  these cases, all the geodesics will be of maximal length compared to nearby (non-extremal) trajectories.

\subsection{Working in dimensionless coordinates} \label{dimlescoord}
To simplify the notation in what follows, we will be using dimensionless coordinates. We redefine the radial coordinate $r_{dl}=r/|r_h|$, the dilaton  potential $U(\phi) = U_{dl}(r_{dl}) \, |r_h| /\ell$ and the blackening factor $f(r) = f_{dl}(r_{dl})\, r_h^2/\ell^2$ such that equation \eqref{blackfactor} becomes 
\begin{equation}
f_{dl}(r_{dl}) = \int_{\text{sign}(r_h)}^{r_{dl}} U_{dl}(r_{dl}) dr_{dl} \,.
\end{equation}
Redefining a dimensionless volume $V_{dl} = V/\ell$, the volume integral \eqref{volume} becomes
\begin{equation}
V_{dl}[P_{dl}] = 2 \int_{r_{dl,t}}^{R_{dl,b}} \frac{dr_{dl} }{\sqrt{f_{dl}(r) + P_{dl}^2}} \,,
\end{equation}
where we have used the redefinitions $r_{dl,t}=r_t/|r_h|$ and $R_{dl,b} = R_b/|r_h|$ and redefined the momentum according to $P = P_{dl} \,|r_h|/\ell$. 
Finally, the time \eqref{time_int}  reads
\begin{equation}
 \frac{t_{dl,R} + t_{dl,L}}{2} = r^*_{dl,t} - r^*_{dl} (R_{dl,b})  +  
\int_{r_{dl,t}}^{R_{dl,b}}  \frac{\sqrt{f_{dl}(r_{dl}) + P_{dl}^2} - P_{dl}}{ f_{dl}(r_{dl})\,  \sqrt{f(r_{dl}) + P_{dl}^2}}  \, dr_{dl} \,,
\end{equation}
where we have redefined the tortoise coordinate \eqref{tortoisy} $r^*(r) = r^*_{dl}(r_{dl})  \, \ell^2/|r_h|$, and the times $t =t_{dl} \, \ell^2/|r_h| $.
Effectively, working in dimensionless conventions simply amounts to setting  $\ell=|r_h|=1$ in all our previous formulas. 
From now on we will do so. We omit the $dl$ subscripts to keep the notation compact and keep in mind that in order to recover the dimensionful volume and time we should substitute
\begin{equation}\label{dimrec}
V= \ell \, V_{dl} , \qquad t =  t_{dl}\, \ell^2/|r_h| \, .
\end{equation}
\section{Geodesics in A(dS) spacetimes} \label{sec_max}

\subsection{Geodesics in global AdS$_2$}

\begin{figure}[h]
	\centering
	\includegraphics[scale=0.5]{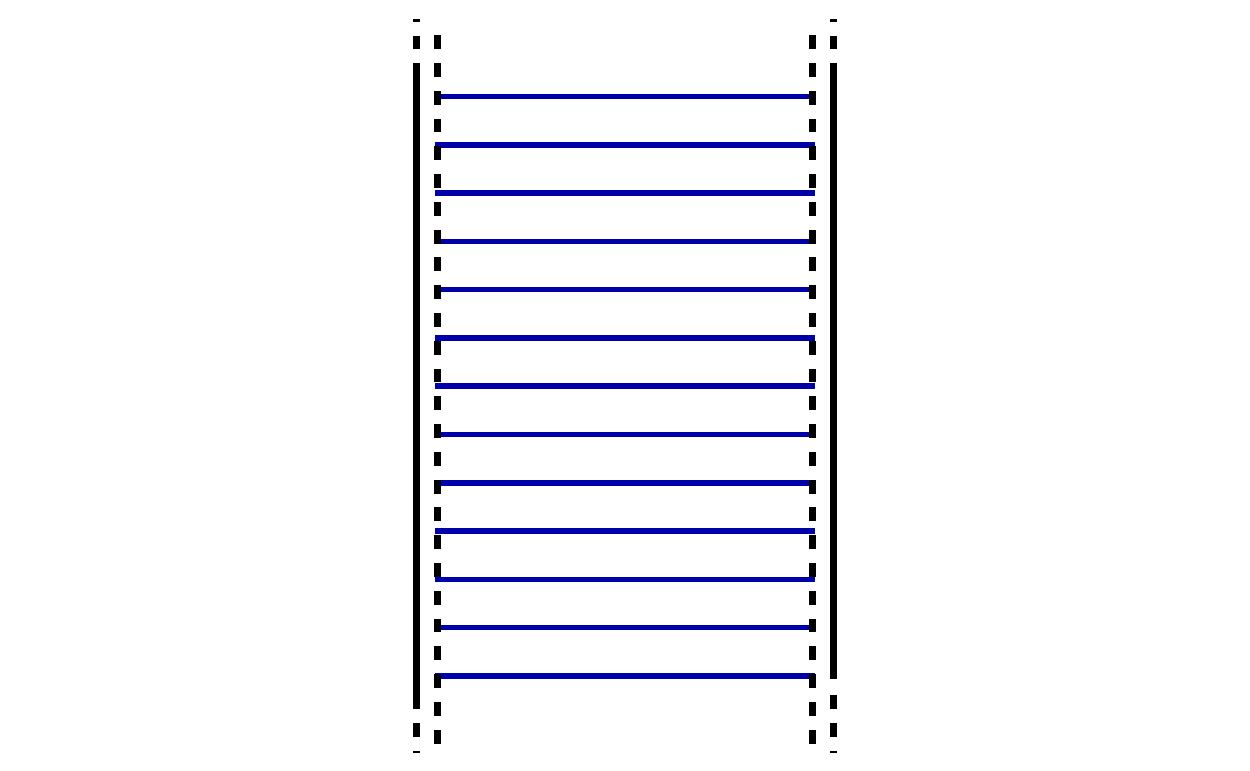}
	\caption{Penrose diagram for global AdS$_2$. The geodesics in blue connect equal times on the two boundaries. The black dashed line is the cutoff surface $r=R_b \gg 1$.}
	\label{fig:global_ads_pen}
\end{figure}

Solutions with constant negative curvature are obtained when $U(\phi) = 2\phi$.\footnote{More generally, the Ricci scalar is give by $R=-f''(r)$.} The metric for global AdS$_2$ is given by\footnote{This geometry, which does not have a horizon, can be obtained by analytically continuing eq.~\eqref{blackfactor} to imaginary horizon radius $r_h = i$.}
\begin{equation}
f(r)_{\text{global}}= r^2+1 \,.
\end{equation}
The Penrose diagram is the infinite vertical strip. It has two boundaries and, in this coordinate system, $r$ runs from $r=-\infty$ at one boundary to $r=+\infty$ at the other. It is known that geodesics at fixed, equal boundary times are just constant global time slices, as shown in figure \ref{fig:global_ads_pen}. So it is straightforward to compute their volume,
\begin{equation}
V = \int^{R_b}_{-R_b} \frac{dr}{\sqrt{f(r)_{\text{global}}}} = \left. \text{arcsinh} \, r \right|^{R_b}_{-R_b} = 2 \log (2 R_b) + O( 1/R_b) \,, \label{global_vol}
\end{equation}
where $R_b$ here serves as a UV regulator for the volume divergences near $|r|=\infty$. 
As expected, the volume is independent of the boundary times chosen. As a warm up exercise, we will use our general $f(r)$ procedure from section \ref{genfrproc} to reproduce this result.

First, we note that in this geometry, geodesics do not have a turning point, \ie $f(r) + P^2$ is always greater than zero, so instead of integrating from the turning point, we just integrate from one boundary to the other. The volume integral \eqref{volume} gives
\begin{equation}\label{volglo1}
V = \int^{R_b}_{-R_b} \frac{dr}{\sqrt{f(r)_{\text{global}} + P^2}} \,.
\end{equation}
To recover eq.~\eqref{global_vol}, we need to show that $P=0$ for geodesics anchored at equal boundary times. 
To demonstrate this, we consider the time integral (\ref{v_eq}) and integrate from one boundary to the other
\begin{equation}
v_R (R_b) - v_R (-R_b) = \int_{-R_b}^{R_b} \tau(P,r) dr \,.
\end{equation}
Using the definition of $v_R$ \eqref{lightcoords}, this can be re-expressed as
\begin{equation}
t_R (R_b) - t_R (-R_b) = r^*(-R_b) - r^*(R_b) + 2 \, \text{arctan} (R_b)-2\, \text{arctan} \left(\frac{P R_b}{\sqrt{P^2+r^2+1}}\right) \, .
\end{equation}
The tortoise coordinate, vanishing at $r\to \infty$, is given by
\begin{equation}
r^*(r) = \text{arctan} \, r - \frac{\pi}{2} \,.
\end{equation}
With these ingredients, the requirement that the geodesics are anchored on both boundaries at the same time $t_R(R_b) = t_R(-R_b)$ yields
\begin{equation}
0 = -2 \, \text{arctan}\left(\frac{P R_b}{\sqrt{P^2+R_b^2+1}}\right)  \,,
\end{equation}
which sets $P=0$. Then our volume integral \eqref{volglo1} reduces to the one found in equation (\ref{global_vol}), as expected.

\subsection{Geodesics in the AdS$_2$ black hole}
The AdS$_2$ black hole is actually very similar to the previous global AdS$_2$. The difference is that the metric is expressed in Rindler coordinates and the boundary, located at some constant value of the Rindler radial coordinate, is bent towards the bulk in a way which makes parts of it inaccessible, \ie hidden behind a horizon, see figure \ref{fig:penroseperlim}.  The complexity of the AdS$_2$ black hole was already studied in \cite{Brown:2018bms}, where it was found that the geodesic length grows linearly with time at late times. This result  is derived using the known fact that geodesics are lines of constant ``global'' time. Here, we  reproduce that behaviour using the procedure described in section \ref{genfrproc}.

The AdS$_2$ black hole is obtained once again using the  dilaton potential $U(\phi) = 2\phi$. The corresponding blackening factor reads
\begin{equation}
f(r)_{\text{BH}} = r^2 -1 \,,
\end{equation}
where we set the horizon and curvature radii $r_h=\ell=1$ as described in section \ref{dimlescoord}. This corresponds to a temperature of $T = 1/2\pi$, see eq.~\eqref{thermoquants}. 
The boundary of AdS is at $r\to \infty$. In higher dimensional black holes there is a curvature singularity at $r \to 0$, but this is not the case in two dimensions. 

\begin{figure}[h!]
        \centering
        \subfigure[]{
                \includegraphics[scale=0.35]{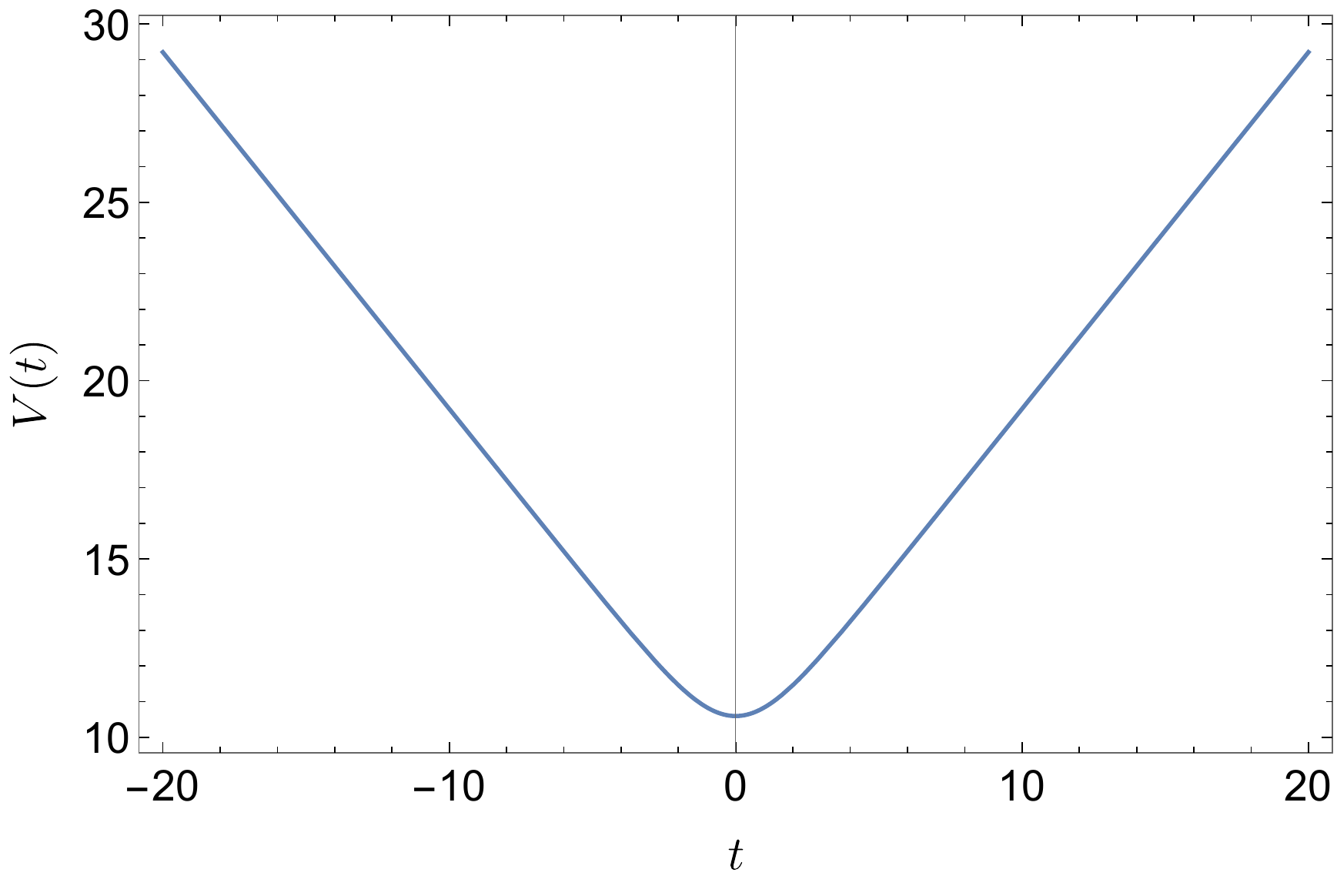} } \quad\quad
         \subfigure[]{
                \includegraphics[scale=0.37]{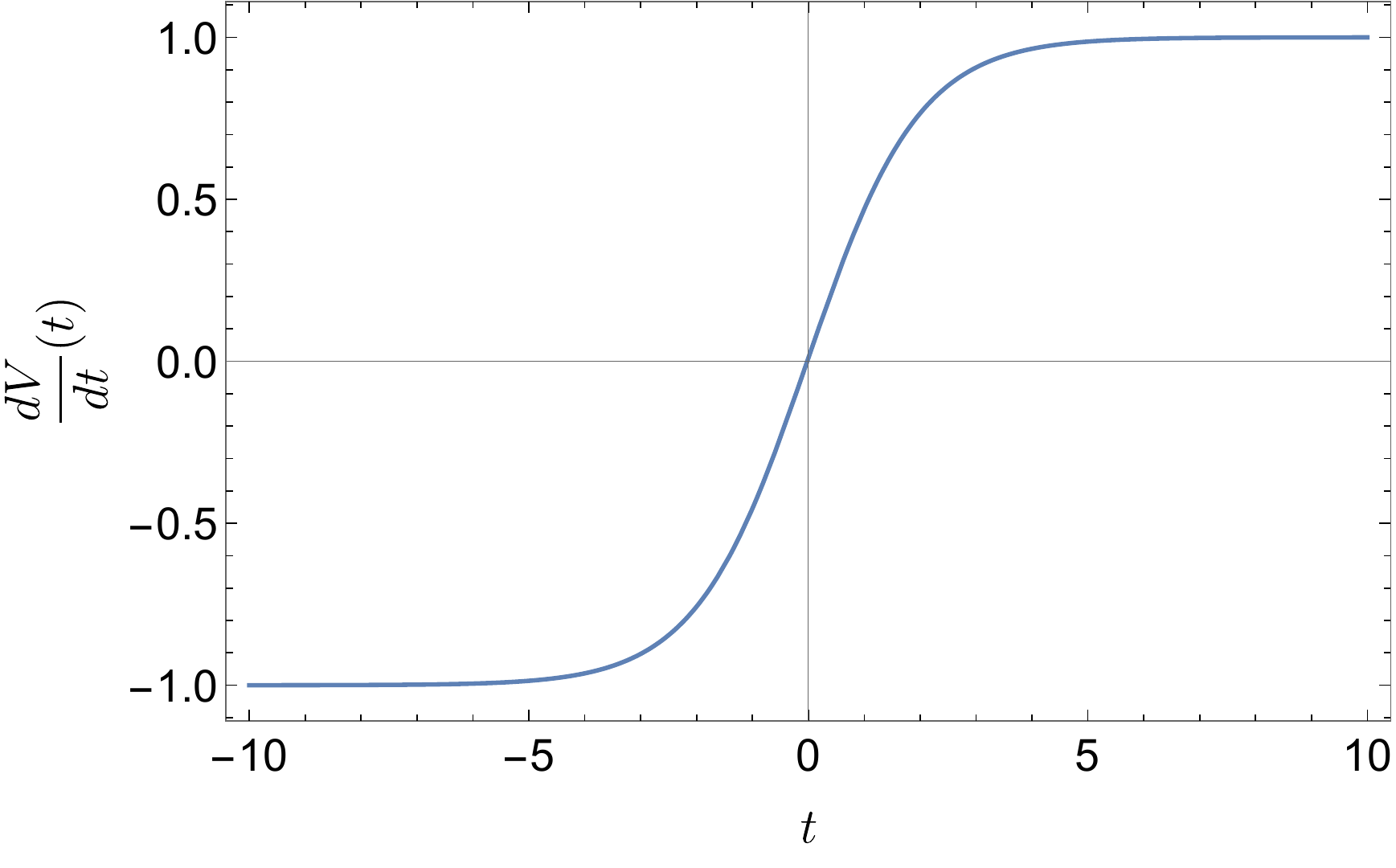} }
                 \caption{{\footnotesize Volume and its time derivative as a function of the boundary time $t$ with $R_b = 100$. }}
\label{fig_ads_bh}
\end{figure}

We can follow the procedure outlined in the previous section. First, we need to find the turning point,
\begin{equation}
f(r_t) + P^2 = r_t^2 - 1 + P^2 = 0 \rightarrow r_t = \sqrt{1-P^2} \,,
\end{equation}
which gives a turning point $r_t \leq 1$ inside the horizon and implies that $-1 \leq P \leq 1$. Next we need to perform the volume and times integrals. 
The volume integral (\ref{volume}) can be performed  analytically and yields 
\begin{equation}
V[P]  =2\,{\rm arccosh}\left(\frac{R_b}{\sqrt{1-P^2}}\right)\,.
\label{vol_ads_bh}
\end{equation} 
The time integral (\ref{time_int}) can also be evaluated   analytically. We first note that the tortoise coordinate $r^*(r) = \frac{1}{2}\log \left|\frac{r-1}{r+1}\right|$ at the turning point is given by
\begin{equation}
r^*_t =  -\text{arccosh}\left(\frac{1}{|P|}\right)\,.
\end{equation}
Then,  \eqref{time_int} becomes 
\begin{equation}
t =2\,{\rm arctanh}\left(\frac{P R_b}{\sqrt{R_b^2-1+P^2}}\right)\,.
\end{equation}
Note that for large $R_b$, this expression gives $t_L + t_R = 0$ for $P=0$ and $t_L  + t_R \to \pm \infty$ for $P=\pm 1$, so it covers all boundary times. Luckily, it is also possible to invert this expression analytically and obtain
\begin{equation}
P = \frac{ \tanh \left(\frac{t}{2}\right) \sqrt{R_b^2-1}}{\sqrt{R_b^2-\tanh ^2\left(\frac{t}{2}\right)}}\, .
\end{equation}
Plugging this into (\ref{vol_ads_bh}) we find that
\begin{equation}
V(t) =2\,{\rm arccosh}\,\left(\sqrt{(R_b^2-1)\cosh^2\frac{t}{2}+1}\right).
\end{equation}
We plot this function in figure \ref{fig_ads_bh}, where a linear growth at late times can be observed. In fact, if we expand this expression for large $R_b$, we obtain
\begin{equation}
V(t) = 2 \log \left(2 R_b \cosh \frac{t}{2} \right) + O(1/R_b^2) \,, \label{vol_bh_t}
\end{equation}
which becomes at late times 
\begin{equation}\label{totvolbhads}
V(t) \approx 2 \log R_b + |t| + \cdots \,,
\end{equation}
which is the celebrated linear growth result. To eliminate the cutoff dependence, we may consider the time derivative of the volume 
\begin{equation}
\frac{dV}{dt} = 
 \frac{ \tanh \left(\frac{t}{2}\right) \sqrt{R_b^2-1}}{\sqrt{R_b^2-\tanh ^2\left(\frac{t}{2}\right)}} = P
= \tanh \frac{t}{2} + O(1/R_b^2) \xrightarrow[t\to \infty]{}  1 \,.
\end{equation}
Recall that the equality to $P$ is a general property of the rate of change of the volume, see comments around equation \eqref{dVdtgen}.
Re-establishing the dimensions using equation \eqref{dimrec} and the thermodynamic quantities  \eqref{thermoquants} we obtain in the late time limit
\begin{equation}
\frac{dV}{dt} = r_h/\ell = 2\pi \ell \, T, \qquad
\frac{d\mathcal{C}_V}{dt} = 8 \pi S_0 T,
\end{equation}
where the complexity was evaluated using eq.~\eqref{CV} with the extra factor of $\Phi_0$ suggested by \cite{Brown:2018bms}, and $S_0$ is the leading contribution to the entropy.

We can instead consider only the part of the volume that lies behind the horizon. This requires integrating \eqref{volume} from the turning point $r_t = \sqrt{1-P^2}$ to the horizon $r_h =1$, giving
\begin{align}
V_{\rm inside} &= 2\,{\rm arccosh}\left(\frac{1}{\sqrt{1-P^2}}\right) \nonumber\\
&= 2\,{\rm arccosh}\left(\frac{\sqrt{(R_b^2-1)\cosh^2\frac{t}{2}+1}}{R_b}\right)
= |t|+O\left(1/R_b^2\right)\;.	
\end{align}
Subtracting this from eq.~\eqref{totvolbhads}, we see that the volume outside the horizon does not grow linearly and in fact approaches a constant at late times.

\begin{figure}[h]
	\centering
	\includegraphics[scale=0.5]{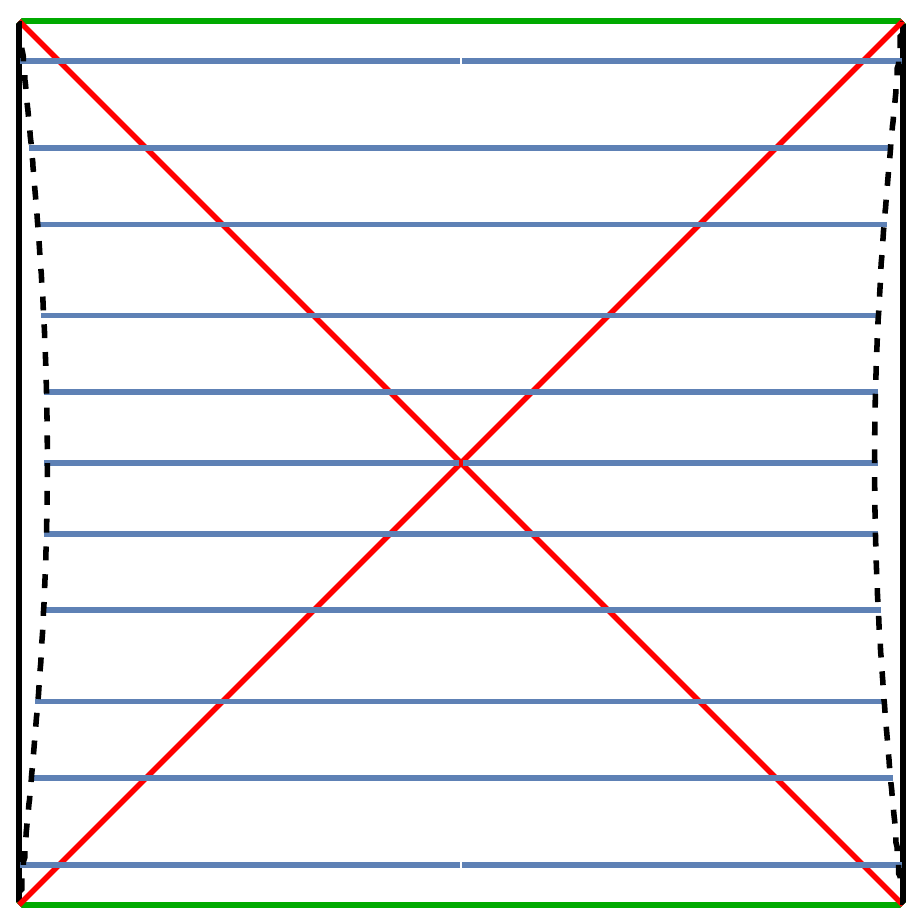}
	\caption{Penrose diagram for the AdS$_2$ black hole and the  geodesics in blue connecting equal times at the two boundaries. $R_b=10$ is the black dashed line.}
	\label{fig:adsbh_pen}
\end{figure}

A collection of geodesics anchored at different boundary times in the AdS$_2$ black hole Penrose diagram is shown in figure \ref{fig:adsbh_pen}. Note that the geodesics are indeed constant global time slices.\footnote{It is interesting to compare this shape of the geodesics to the extremal volumes obtained for higher dimensional black holes. In the latter case, the extremal slices wrap around constant $r_{\text{min}}=r_h/2^{1/d}$ at late times, see section 3.1 of \cite{Carmi:2017jqz}. This is obtained by minimizing the turning point associated function  $W(r) = \sqrt{-f(r)} r^{d-1}$, but since for us $d=1$ the minimization yields $r_{\text{min}}=0$ and all the slices approach this straight line.}

\subsection{Geodesics in dS$_2$} \label{sec_ds}

In order to obtain solutions with positive constant curvature,  we set $U(\phi) = -2\phi$.  The metric corresponds to pure dS$_2$, with a blackening factor given by
\begin{equation}
f(r)_{\text{dS}} = 1- r^2 \,.
\end{equation}
We will concentrate on the part of the geometry with positive $r$. The radial coordinate outside the horizon ranges  between $0 \leq r \leq 1$ and future/past infinity is reached as $r \to \infty$.

Clearly, in this geometry there is no timelike boundary where it is natural to anchor the geodesics. Nonetheless, we can consider symmetric geodesics anchored at the observer's worldline $r=0$. This will provide some interesting intuition for the geodesics in the flow geometries which we consider in the next section. The technology is very similar to that developed in section \ref{genfrproc}. 
Geodesics in the dS$_2$ spacetime have a turning point at
\begin{equation}
r_t = \sqrt{1 + P^2} \,,
\end{equation}
and the tortoise coordinate is 
\begin{equation}
r^* (r ) = \frac{1}{2} \log \left| \frac{r+1}{r-1 } \right| \,,
\end{equation}
with the integration constant chosen so that the tortoise coordinate vanishes on the  observer's worldline $r^* (r=0) = 0$. At the turning point this results in 
\begin{equation}
r_t^* ={\rm arcsinh}\,\left(\frac{1}{|P|}\right)\,.
\end{equation}

\begin{figure}[h]
	\centering
	\includegraphics[scale=0.5]{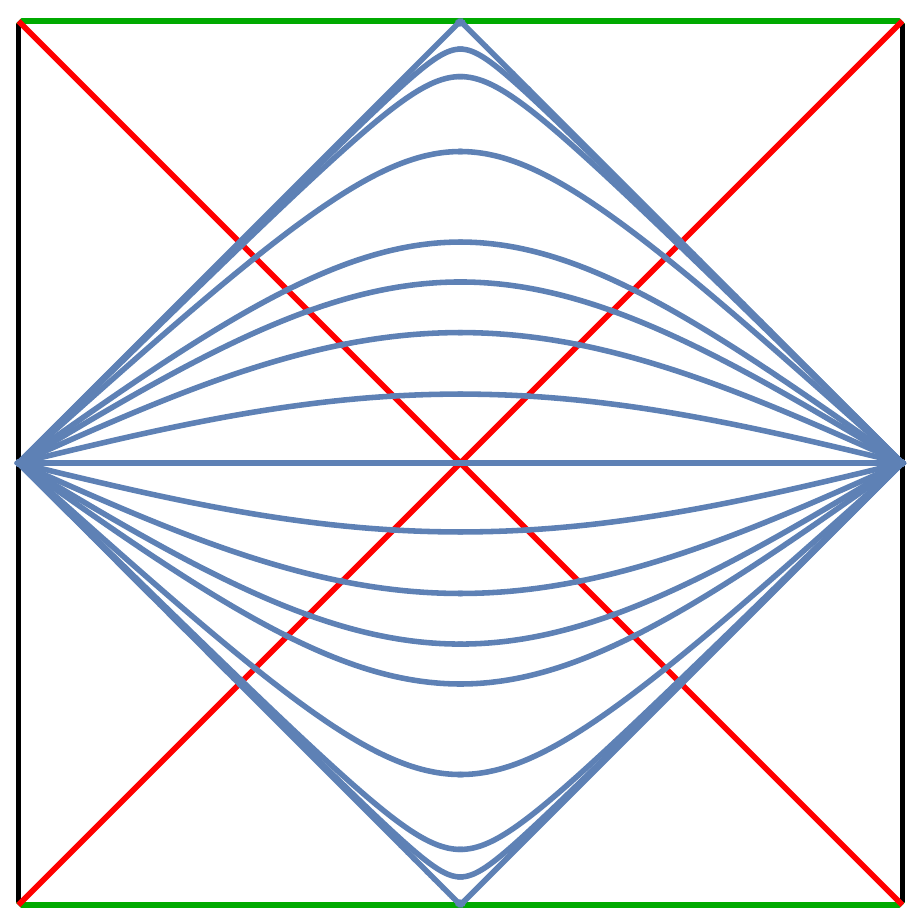}
	\caption{Penrose diagram for (half of) dS$_2$ and the geodesics in blue connecting $t_R=t_L=0$. $P$ runs from $-\infty$ to $\infty$ and in these limits, the geodesics become (almost everywhere) null.}
	\label{fig:ds_pen}
\end{figure}

The volume integral \eqref{volume} does not depend on $P$ and in fact, we obtain
\begin{equation}
V [P] = \pi \,,
\end{equation}
for any spacelike geodesic in dS anchored at points with $r=0$. Evaluating the time integral, we obtain 
\begin{equation}
\frac{t_R + t_L}{2} = r^*_t  + \int_{r_t}^{0} \tau (P,r) dr = 0 \,.
\end{equation}
We see that geodesics anchored on the left and right $r=0$ worldlines must satisfy  $t_R = - t_L$. 
This means that certain points on the worldlines are not connected by smooth spacelike geodesics of finite length. The Penrose diagram with the symmetric $t_L=t_R=0$ geodesics are shown in figure \ref{fig:ds_pen}. 
We did not consider geodesics reaching future/past infinity whose lengths diverge. We will return to this point in section \ref{discussion}.

\section{Geodesics in flow geometries} \label{sec_flow}
\subsection{Geodesics in the centaur geometry}
With our experience of AdS$_2$ and dS$_2$, we can now address the problem of finding geodesics in the centaur geometry
\begin{align}
f(r)_{\text{centaur}} &= \begin{cases}
	(1- r^2) \,, & -\infty < r < 0 \,,
\\
	(1+ r^2) \,, & 0 < r < \infty \,.
	\end{cases}
\end{align}
This geometry is obtained as a solution of a dilaton-gravity theory with potential $U(\phi) = 2 |\phi|$.\footnote{One might worry that this potential has a discontinuity in its derivative. It has been shown in \cite{Anninos:2017hhn} that this is not a problem, since it can be thought of as a smooth limit of continuous dilaton potentials.} 
For large $r\to\infty$, the geometry looks like global AdS$_2$. At $r=0$, it interpolates into a dS$_2$ region with a horizon at $r=-1$. The choice of the horizon\footnote{Recall that in two dimensions there are two cosmological horizons at $r_h=\pm 1$. } at $r_h = -1$ was made such that it gives the same temperature as for the black hole case $T = {1}/{2\pi}$, see Appendix \ref{app_thermo}.
We refer to this geometry as a centaur geometry.

\begin{figure}[h!]
        \centering
        \subfigure[$P=-1$.]{
                \includegraphics[scale=0.6]{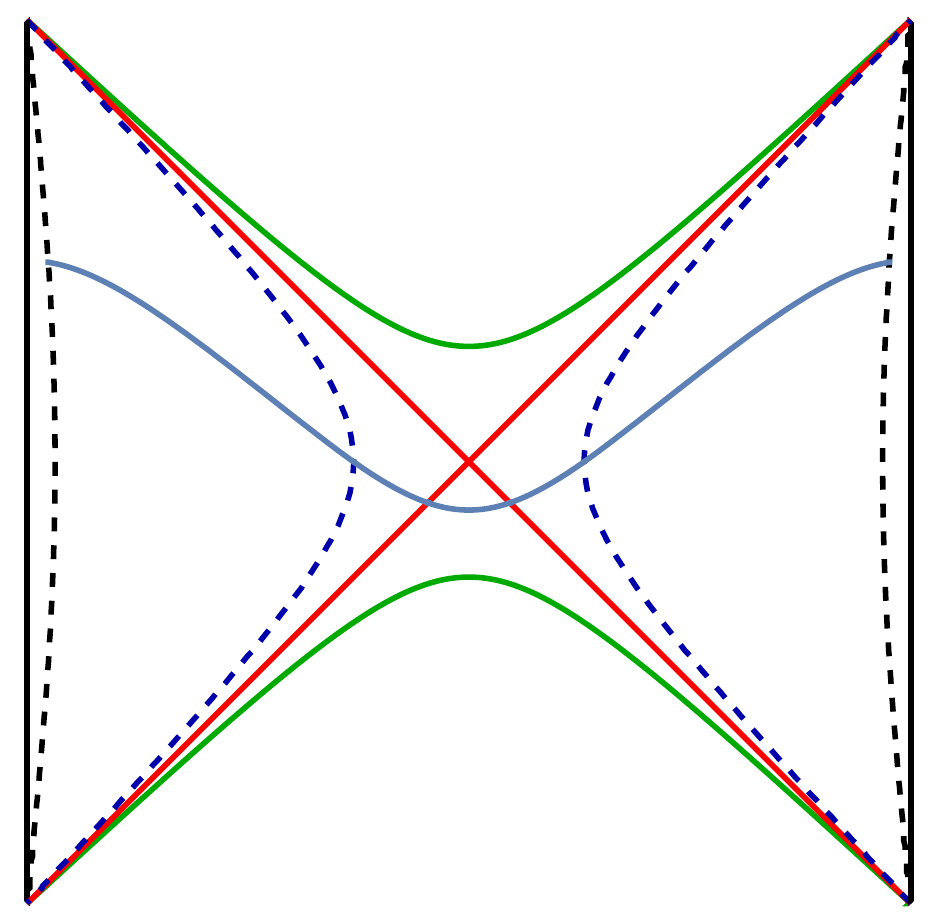} \label{centaurP1}} \quad\quad
         \subfigure[$P= \pm (100, 2, 1, 0.5, 0.25, 0.01)$.]{
                \includegraphics[scale=0.6]{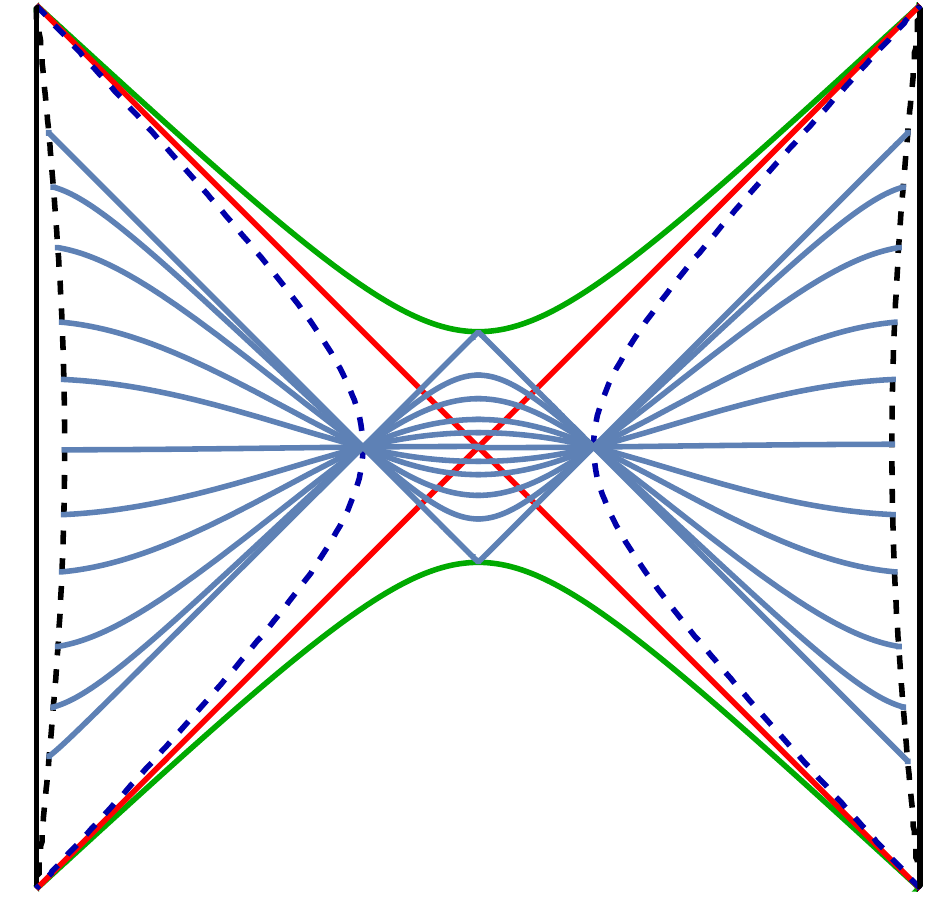} }
                 \caption{\footnotesize Penrose diagrams for the centaur geometry and geodesics anchored at different boundary times in blue, spanning the full range of times for which smooth spacelike geodesics of finite length exist. The dashed black line is the cutoff surface with $R_b=10$. The dark blue dashed line is the interpolating line between the two geometries at $r=0$. The red lines correspond to the horizons and the green ones correspond to $r\to -\infty$.}
\label{fig:centaur_pen}
\end{figure}

We assume there is a turning point $r_t$ along our geodesics with
\begin{equation}
r_t = -\sqrt{1+P^2} \,.
\end{equation}
Note that $r_t<0$, which implies that the turning point is inside the dS horizon. The next step is to define the tortoise coordinate. Again, we fix $r^*(r)$ so that it vanishes at the boundary and we require continuity along the interpolating curve $r=0$. Doing so, we obtain,
\begin{align}
r^*(r)_{\text{centaur}} &= \begin{cases}
	\frac{1}{2} \log \left| \frac{ r+1 }{r-1 }\right|-\frac{\pi }{2}  \,, & -\infty < r < 0 \,,
\\
	\text{arctan} (r)-\frac{\pi }{2} \,, & 0 < r < \infty \,.
	\end{cases}
\end{align}
Evaluating this at the turning point we further get 
\begin{equation}
r^*_t = -\frac{\pi}{2}-{\rm arcsinh}\left(\frac{1}{|P|}\right)\,.
\end{equation}
We can evaluate the volume and time integrals separating the integrals into intervals. Note that since the metric is continuous up to its first derivatives, there is no jump in $P$ along the geodesic.\footnote{This can be proven using the equations of motion for the volume \eqref{volume2} integrated in a small shell  around the transition between the dS and the AdS regions.}   
The volume integral (\ref{volume}) yields
\begin{equation}
\begin{split}
V[P] & =  2 \left(\int_{\text{dS}} + \int_{\text{AdS}} \right) \frac{dr}{\sqrt{f(r)_{\text{centaur}} + P^2}}\\
& =  2 \left( \int_{r_t}^{0} \frac{dr}{\sqrt{1 - r^2 + P^2}} +  \int_{0}^{R_b} \frac{dr}{\sqrt{1 + r^2 + P^2}} \right)  \\
& =  \pi + 2 \, \text{arcsinh} \left(\frac{R_b}{\sqrt{P^2+1}}\right) \,. \label{volume_centaur}
\end{split}
\end{equation}
Note that the $\pi$ contribution comes precisely from the dS part and it is the same that we got in section \ref{sec_ds}. The second term is the contribution from the AdS patch. Similarly, we can perform the time integral,
\begin{equation}
\begin{split}
\frac{t}{2} & =  r^*_t  - r^*(R_b) + \int_{r_t}^0 \tau_{\text{dS}} (P,r) dr + \int_0^{R_b} \tau_{\text{AdS}} (P,r) dr \\
& =  r^*_t - r^*(R_b) + \left. \frac{1}{2} \log \left( \left| \frac{(r+1) \left(\sqrt{P^2-r^2+1}-P r\right)}{(r-1) \left(P r+\sqrt{P^2-r^2+1}\right)}\right| \right) \right|^0_{r_t}   \\
&  + \left( \left.\text{arctan} (r)- \text{arctan} \left(\frac{P r}{\sqrt{P^2+r^2+1}}\right) \right) \right|^{R_b}_0  \\
& =  - \text{arctan} \left(\frac{P R_b}{\sqrt{R_b^2+P^2+1}}\right) = - \, \text{arctan} \, P+O\left(1/R_b^2\right) \,. \label{t_centaur}
\end{split}
\end{equation}
It is interesting to note that for positive $P$, the times are negative (and vice versa) and also that while $-\infty < P < \infty$, the times are constrained to the range $-\pi< t < \pi$, so it is not possible to obtain geodesics connecting boundary points at equal arbitrarily large times. This can be appreciated in the Penrose diagram in figure \ref{fig:centaur_pen} where we plot some geodesics.

The reason for this is that in the dS part, the only allowed geodesics start at $t=0$ and not all boundary points in the AdS part are spacelike connected to this point. It would be interesting to understand this intriguing feature of the geometry from the boundary quantum perspective. We return to this point in section \ref{discussion}.

Another important difference between the centaur geometry and the AdS black hole is that for the centaur, geodesics anchored at positive boundary times pass through the past horizon, and not the future one, as can be seen from figure \ref{centaurP1}.

Finally, we can express the volume in terms of the boundary times. For this, we first need to invert equation (\ref{t_centaur}) which yields
\begin{equation}
P^2 = \frac{\left(R_b^2+1\right) \tan ^2\left(\frac{t}{2}\right)}{R_b^2-\tan ^2\left(\frac{t}{2}\right)} \,,
\end{equation}
as long as $R_b >\tan \frac{|t|}{2} $. 
Inserting this into equation (\ref{volume_centaur}) we get
\begin{equation}
V( t) = \pi + 2 \,\text{arcsinh} \left(\sqrt{(R_b^2 +1)\cos^2 \left(\frac{t}{2}\right)-1}\,\right),
\end{equation}
and at large $R_b$ this becomes
\begin{equation}
V (t) = \pi + 2 \log \left(2 R_b \cos \frac{t}{2}\right) + O(1/R_b^2) \,, \label{vol_t_cent}
\end{equation}
which is valid as long as $-\pi < t < \pi $ and $R_b + \tan \frac{|t|}{2} >0$. It is straightforward to compute the time derivative,
\begin{equation}
\frac{dV(t)}{dt} 
=P=-\frac{\sqrt{R_b^2+1} }{\sqrt{R_b^2-\tan ^2\left(\frac{t}{2}\right)}} \tan \left(\frac{t}{2}\right) = -\tan \frac{t}{2}+O\left(1/R_b^2\right) \,.
\end{equation}

Re-establishing the dimensions using eq.~\eqref{dimrec} and the thermodynamic quantities  \eqref{thermoquants} we obtain obtain 
\begin{equation}
\frac{dV}{dt} = -2\pi \ell \,T \tan (\pi t T),\qquad
\frac{d\mathcal{C}_V}{dt} =- 8 \pi S_0 T \tan (\pi t T),
\end{equation}
where the complexity was evaluated using eq.~\eqref{CV} with the extra factor of $\Phi_0$ suggested by \cite{Brown:2018bms}, and $S_0$ is the leading contribution to the entropy.

\begin{figure}[h!]
        \centering
        \subfigure[]{
                \includegraphics[scale=0.5]{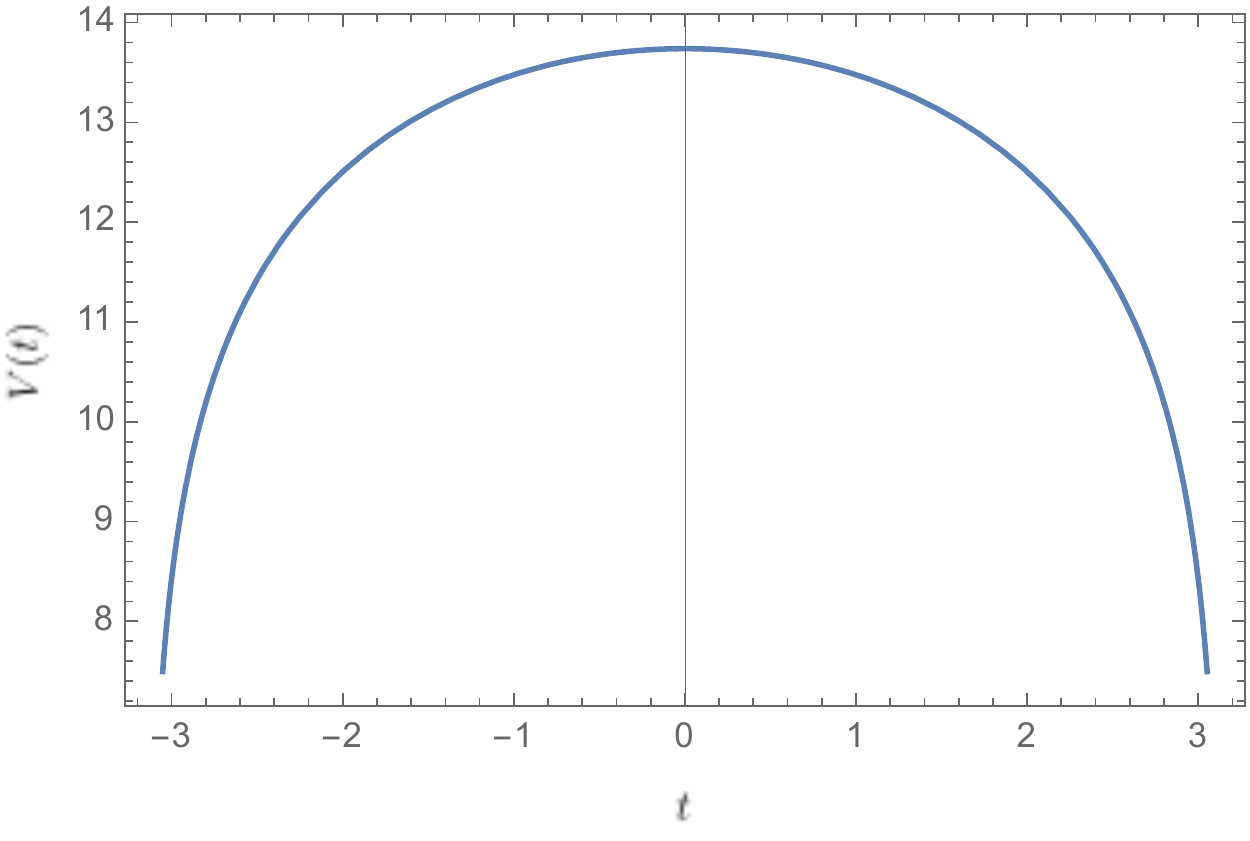} } \quad\quad
         \subfigure[]{
                \includegraphics[scale=0.5]{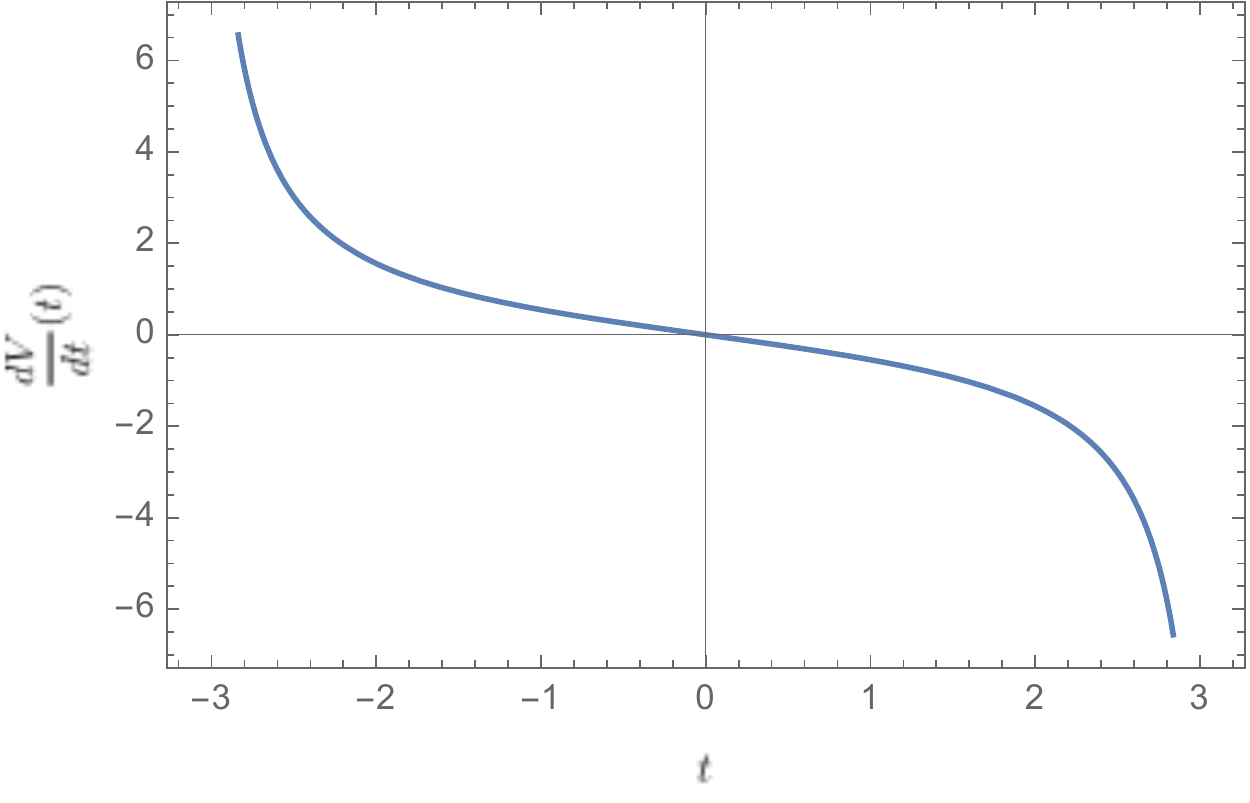} }
                 \caption{{\footnotesize Volume and its time derivative as a function of the boundary time $t$ for the centaur geometry with $R_b = 100$. }}
\label{fig_centaur_V}
\end{figure}

Plots of these functions can be found in figure \ref{fig_centaur_V}. The behaviour exhibited by the centaur geometry is radically different from the one observed in the black hole case, even though both geometries have an event horizon. Comparing  equations (\ref{vol_bh_t}) and  (\ref{vol_t_cent}), we note that they are related by changing $t \to i t$. Nevertheless, their behaviour is completely different. The centaur geometry does not exhibit linear growth of the volume as a function of time and in fact, there is no growth at all but a decrease in volume as time advances. While the the time derivative of the volume in the black hole case goes to a constant, here it diverges when approaching the edges of the range of allowed times. Moreover, the length of the geodesic in the dS part of the geometry remains constant at a value of $\pi$. Finally, this behaviour is not valid for arbitrary long times. After a certain time, there are no connected geodesics between the two boundaries of spacetime.

It is instructive to compute the part of the volume that lies behind the horizon. This requires integrating from the turning point $r_t=-\sqrt{1+P^2}$ to the horizon $r_h=-1$. We find:
\begin{align}
	V_{\rm inside}[P] = & 2  \int_{-\sqrt{1+P^2}}^{-1} \frac{dr}{\sqrt{1 - r^2 + P^2}} = \pi - 2\arcsin\left(\frac{1}{\sqrt{1+P^2}}\right)\;.
\end{align}
Since the geodesic is still anchored to the boundary, we can still use \eqref{t_centaur} to associate a time to it. This yields a very short regime of linear growth
\begin{align}
\begin{split}
\label{volume_inside_centaur}
V_{\rm inside}
	&=\pi - 2\arcsin\left(\frac{\sqrt{(1+R_b^2)\cos^2\frac{t}{2}-1}}{R_b}\right)
	= |t|+O\left({1}/{R_b^2}\right)\;,
\end{split}
\end{align}
valid for times $|t|<\pi$, as follows from  \eqref{t_centaur} with $-\infty <P<\infty$. In contrast, in this range of times, the geodesic length inside the black hole grows quadratically with time, as can be seen from expanding equation (\ref{vol_bh_t}).

\subsection{Geodesics in $\gamma$-centaurs} 
\label{gamma_cen_sec}
Next, we consider a one-parameter family generalization of the centaur geometry from the previous section, characterized by a parameter $\gamma \in \left[-1,1\right] $. These geometries have been proposed in Appendix D of \cite{Anninos:2018svg}. They can be thought of as solutions of a different dilaton-gravity theory with potential $U(\phi) = 2 (|\phi-\phi_0|-\phi_0)$. The parameters $\phi_0$ and $\gamma$ are related according to
\begin{equation}
\gamma \equiv 1 - 2 \phi_0^2 \,.
\end{equation} 
The thermodynamic properties of these theories are explored in Appendix \ref{app_thermo}. 

In the Schwarzschild gauge which we will be using, the metric takes the following form\footnote{While $\phi_0$ is a natural variable in the Schwarzschild gauge, in the conformal gauge it is more convenient to use $\gamma$. The relation between the two coordinate systems is given in Appendix \ref{app_coordinates}.}
\begin{align}
f(r)_{\gamma} &= \begin{cases}
	1- r^2 \,, & -\infty < r < \phi_0 \,,
\\
	1+ r^2+2 \phi _0 \left(\phi _0-2 r\right)  \,, & \phi_0 < r < \infty \,,
	\end{cases}
\end{align}
where we again set $r_h = -1$, to keep the same temperature as before. For $r<\phi_0$, the geometry has positive curvature while for $r>\phi_0$, the curvature is negative. The parameters $\phi_0$ and $\gamma$ can lie in the following ranges
\begin{align}
 \left\{
 \begin{array}{lcl}
 \frac{1}{\sqrt{2}} > \phi_0 > 0 &\rightarrow& 0<\gamma<1 \,, \\
0 > \phi_0 > -\frac{1}{\sqrt{2}}  &\rightarrow& 1 > \gamma > 0 	\,, \\
-\frac{1}{\sqrt{2}} > \phi_0 > -1 &\rightarrow &	 0 > \gamma >-1 \,,
\end{array}
	 	\right.
\end{align}
where $\phi_0 = 0$ corresponds to $\gamma = 1$ and it is the centaur geometry, and when $\gamma = -1$, the interpolating region sticks to the horizon, so from the outside there is only negatively-curved spacetime. Positive $\phi_0$ solutions correspond to having a larger part of the dS spacetime while negative $\phi_0$ corresponds to having a smaller dS portion. 

\begin{figure}[h!]
        \centering
        \subfigure[$\phi_0= 0.2$ ]{
                \includegraphics[scale=0.33]{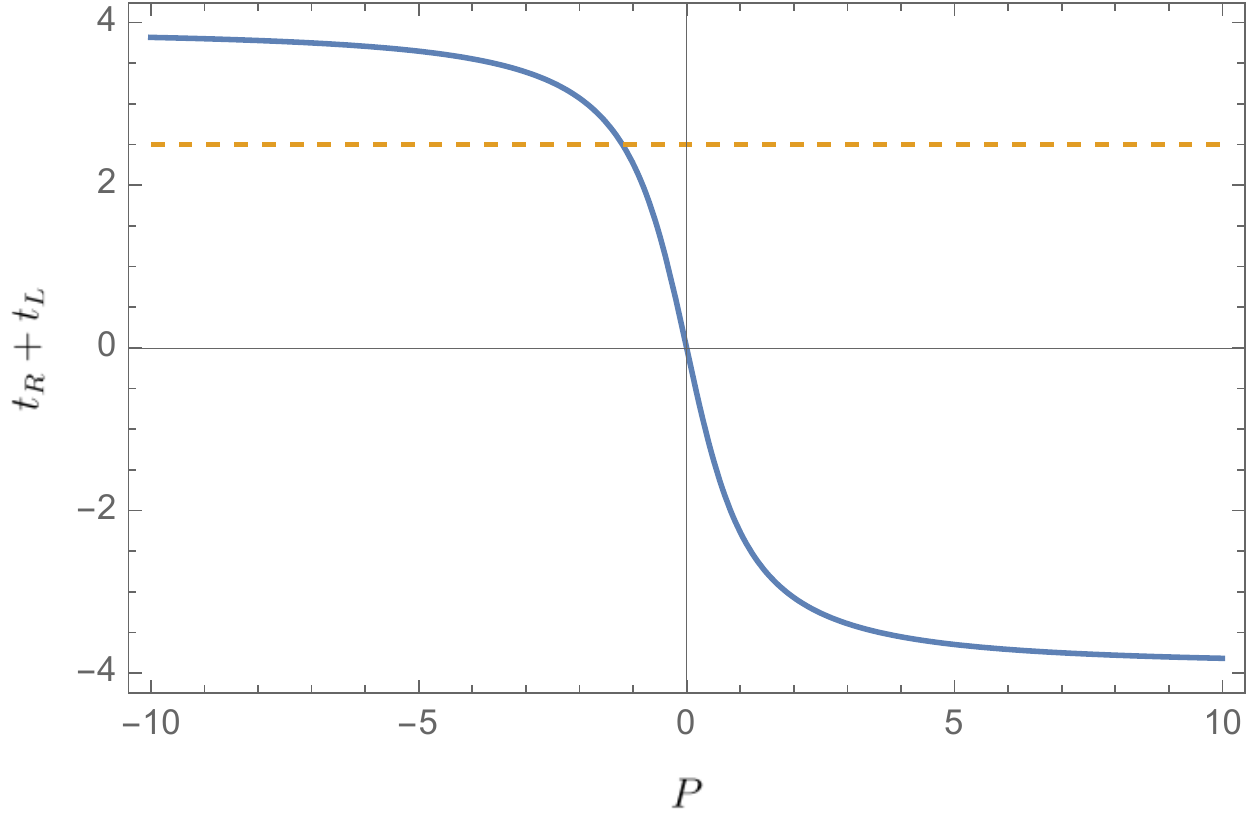} \label{t_gamma_pos}} \quad\quad
                   \subfigure[$\phi_0= -0.85$]{
                \includegraphics[scale=0.33]{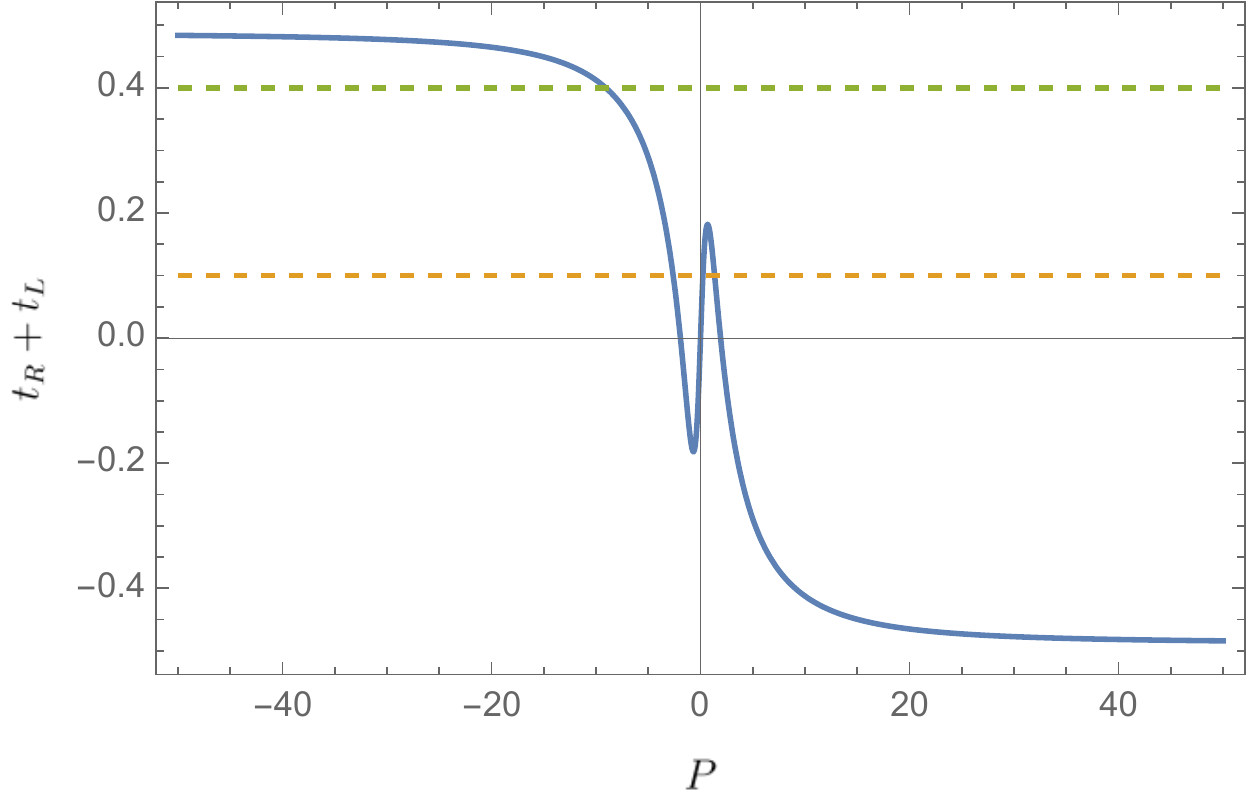} \label{t_gamma_neg1} } \quad\quad
         \subfigure[$\phi_0= -0.95$]{
                \includegraphics[scale=0.33]{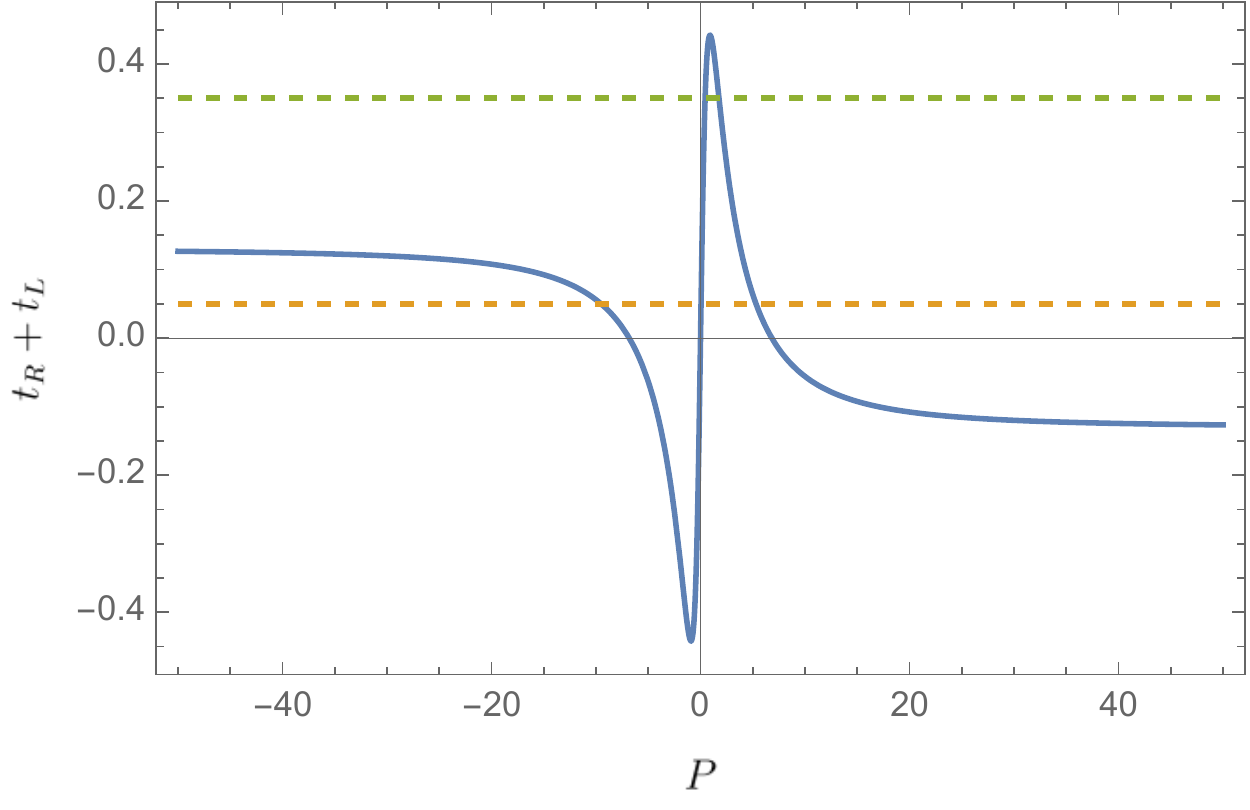} \label{t_gamma_neg2} }
                 \caption{\footnotesize $t_L+t_R$ as a function of $P$ for different values of $\phi_0$. $R_b$ is set to 10. In the first plot, each time corresponds to at most a single value of $P$. In the second plot, at short times there are three relevant values of $P$  (yellow dashed line), while at later times, there is only one relevant value of $P$ (green dashed line). In the last plot, the yellow dashed line shows again three relevant values of $P$, but the green line intersects at two different values of $P$.}
\label{fig_tRPtL}
\end{figure}

Having the form of $f(r)$, we can now proceed to evaluate the different integrals. Luckily, they can be done analytically, even though some of the expressions are quite cumbersome. As can be observed, the metric in the dS part, is the same as in previous examples. Therefore, the turning point is also at the same location
\begin{equation}
r_t = -\sqrt{1+P^2} \,.
\end{equation}
The tortoise coordinate will change slightly, as the interpolation curve changes from $r=0$ to $r=\phi_0$. We will write the formulas for the $\phi_0>0$ case, though they can be extended for $\phi_0<0$. The tortoise coordinate satisfying $r^*(r\rightarrow\infty) = 0$  is given by
\begin{align}
r^*(r)_{\gamma} &= \begin{cases}
	\frac{1}{2} \log \left| \frac{ r+1 }{r-1 }\right| + c_2  \,, & -\infty < r < \phi_0 \,,
\\
	\frac{1}{\sqrt{1-2 \phi _0^2}} \left( \text{arctan} \left( \frac{r-2 \phi _0}{\sqrt{1-2 \phi _0^2}}\right) - \pi/2 \right) \,, & \phi_0 < r < \infty \,,
	\end{cases}
\end{align} 
where the constant $c_2$ is chosen so that the coordinate is continuous along $\phi_0$
\begin{equation}
c_2 = -{\rm arctanh}\,\phi _0 - \frac{1}{\sqrt{1-2 \phi _0^2}} \left( \text{arctan} \left(\frac{\phi _0}{\sqrt{1-2 \phi _0^2}}\right)+\pi/2 \right) \,.
\end{equation}
Evaluating it at the turning point, we obtain,
\begin{equation}
r_t^* =  -{\rm arctanh}\,\phi _0  -\frac{1}{\sqrt{1-2 \phi _0^2}} \left( \text{arctan} \left(\frac{\phi _0}{\sqrt{1-2 \phi _0^2}}\right)+\pi/2 \right) -{\rm \,arcsinh}\!\left(\frac{1}{|P|}\right)\,.
\end{equation}
Note how $\gamma$ naturally appears in the expressions. The volume as a function of $P$ reads
\begin{equation}
\begin{split}\label{volume_gamma}
V[P] & =   \pi + 2\, \text{arctan} \left(\frac{\phi _0}{\sqrt{P^2-\phi _0^2+1}}\right) + 2 \log \left( \frac{R_b-2 \phi _0 + \sqrt{1+ P^2 +R_b^2 -4 \phi _0 R_b+2 \phi _0^2} }{\sqrt{P^2-\phi _0^2+1}-\phi _0}\right) \,  
\\
& =  \pi + 2 \log \, 2 R_b +2 \, \text{arctan} \left(\frac{\phi _0}{\sqrt{1+P^2-\phi _0^2}}\right)  -2 \log \left(\sqrt{1+P^2-\phi _0^2}-\phi _0\right)+ O (1/R_b) \,.  
\end{split}
\end{equation}
%
The expression for the time as a function of $P$ is more complicated. For $\phi_0>-1/\sqrt{2}$, we obtain\footnote{For $\phi_0<-1/\sqrt{2}$, part of the expression becomes complex. In order to obtain the right value for the boundary times, one has to take the real part of this expression.}
 \begin{equation}
t \equiv t_L+t_R  =  \log \left(\frac{\sqrt{P^2-\phi _0^2+1}-P \phi _0}{\sqrt{P^2-\phi _0^2+1}+P \phi _0}\right)+ \frac{\text{arctan}_2(x_1, y_1) + \text{arctan}_2(x_2, y_2)}{\sqrt{1-2 \phi _0^2}}   \,,      \label{time_gamma}  
 \end{equation}
where the function arctan$_2(x,y)$ gives the arctangent of $y/x$, taking into account which quadrant the point $(x,y)$ is in and
\begin{equation}
\begin{split}
x_1 & \equiv  -3 \left(P^2+1\right) \phi _0^2+P^2+2 \phi _0^4+1 \,,\\ 
y_1 & \equiv  -2 P \phi _0 \sqrt{1-2 \phi _0^2} \sqrt{P^2-\phi _0^2+1}  \,,\\
x_2 & \equiv  -P^2 \left(-4 \phi _0 R_b+R_b^2+6 \phi _0^2-1\right) + \left(2 \phi _0^2-1\right) \left(-4 \phi _0 R_b+R_b^2+2 \phi _0^2+1\right) ,\\
y_2 & \equiv  2 P \sqrt{1-2 \phi _0^2} \left(2 \phi _0-R_b\right) \sqrt{-4 \phi _0 R_b+R_b^2+P^2+2 \phi _0^2+1} \,.
\end{split}
\end{equation} 
%
\newline\newline
{\bf Behaviour of the boundary times.} Plotting the boundary time for different possible values of $\phi_0$ allows us to see different interesting patterns for the geodesics. For $-\frac{1}{\sqrt{2}}<\phi_0<\frac{1}{\sqrt{2}}$ the behaviour is similar to that of the centaur geometry in the previous section, where each time has at most one specific value of $P$ associated with it, see figure \ref{t_gamma_pos}. Recall that geodesics do not exist for all boundary times. In fact, it can be shown analytically that, for 
$0<\phi_0<\frac{1}{\sqrt{2}}$ the range of times where geodesics exist is 
\begin{equation}
|t| \leq \frac{1}{\sqrt{1-2 \phi _0^2}} \left(\pi + 2 \, \text{arctan} \left(\frac{2 \phi _0 \sqrt{1-2 \phi _0^2}}{-3 \phi _0^2+\sqrt{\left(\phi _0^2-1\right){}^2}+1}\right) \right) - \log \left(\frac{1-\phi _0}{1+ \phi _0}\right) + O (1/R_b)\,,
\end{equation}
where, of course, for $\phi_0 = 0$, this gives the range $|t|<\pi$ as in the previous section. A similar expression can be written in the range $-\frac{1}{\sqrt{2}}<\phi_0<0$. Moreover, as $|\phi_0| \to \frac{1}{\sqrt{2}}$, the time gap goes to infinity, allowing geodesics at all times. In this case the $\gamma\rightarrow 0^+$ centaur geometry develops an infinitely long AdS throat, see \citep{Anninos:2018svg}.

\begin{figure}[h!]
        \centering
        \subfigure[$\phi_0= 0.2$ ]{
                \includegraphics[scale=0.4]{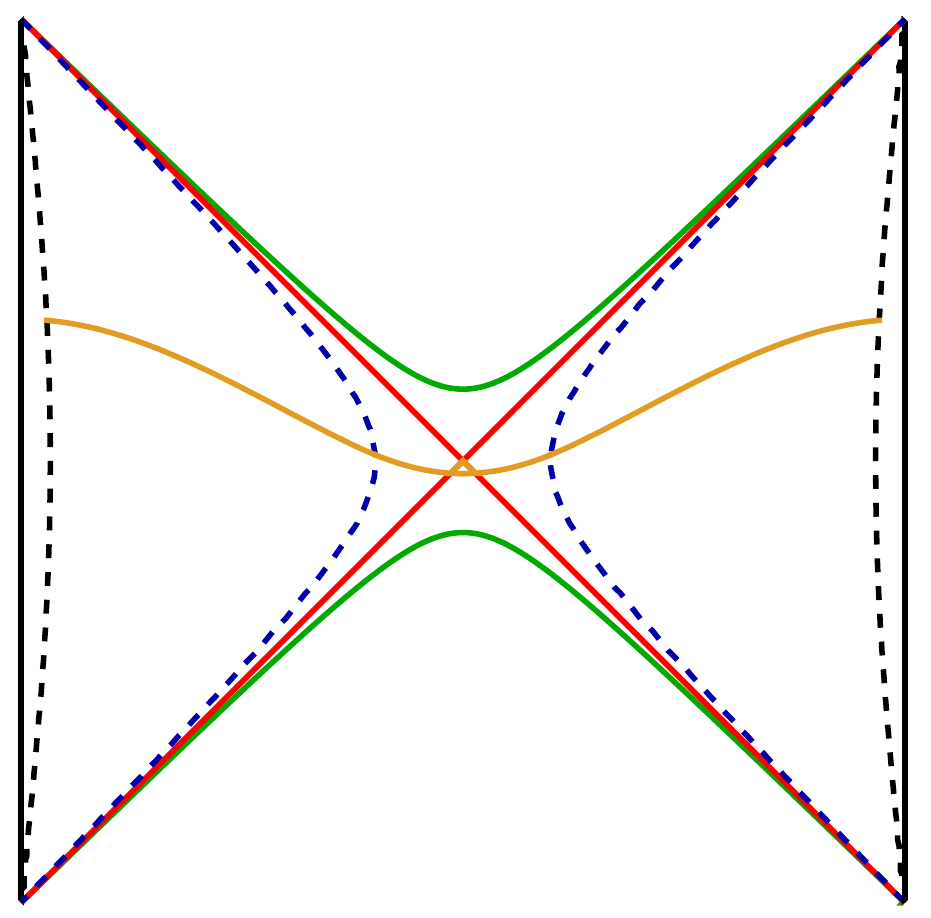} \label{pen_gamma_pos}} \quad\quad
                   \subfigure[$\phi_0= -0.85$]{
                \includegraphics[scale=0.4]{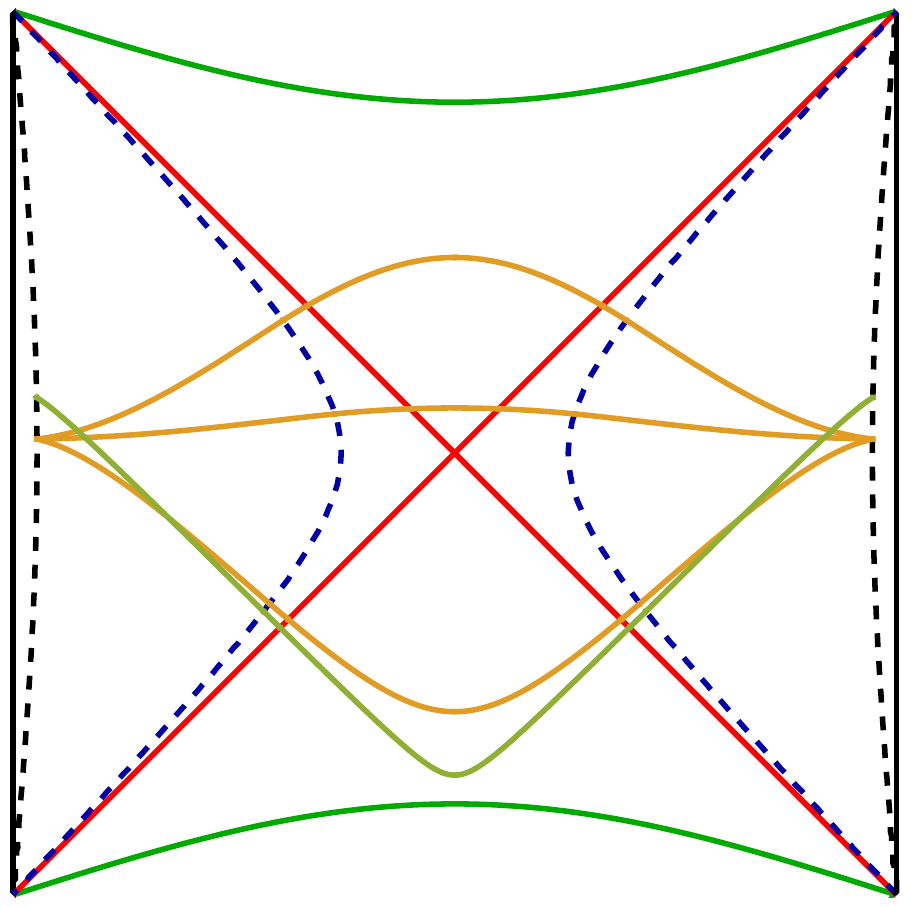} \label{pen_gamma_neg1} } \quad\quad
         \subfigure[$\phi_0= -0.95$]{
                \includegraphics[scale=0.4]{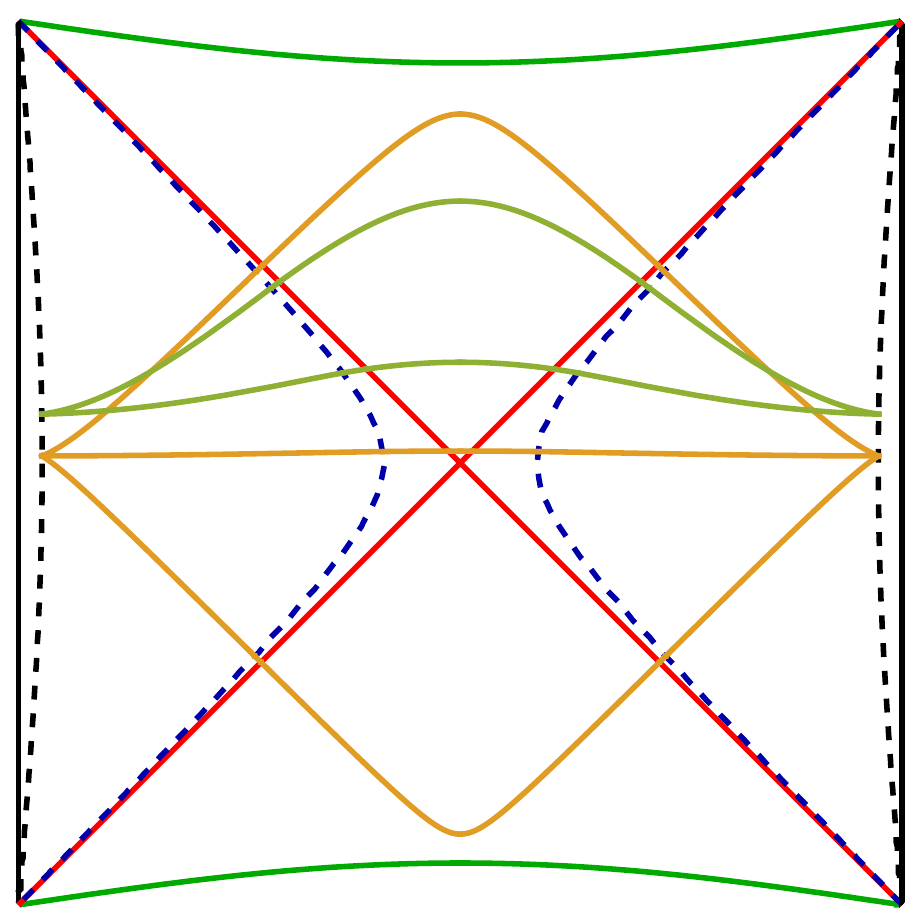} \label{pen_gamma_neg2} }
                 \caption{{\footnotesize Penrose diagrams for the different $\gamma$-centaur geometries and geodesics anchored at different boundary times. In each figure, the colours of the geodesics correspond to those same colours of the dashed lines in figure \ref{fig_tRPtL}. The dashed black line is the cutoff surface with $R_b=10$. The dark blue dashed line is the interpolating line at $r=\phi_0$, the red lines are the horizons and the green ones are $r\to -\infty$. }}
\label{fig_pen_gamma}
\end{figure}

\begin{figure}[h!]
        \centering
        \subfigure[$\phi_0= 0.2$ ]{
                \includegraphics[scale=0.4]{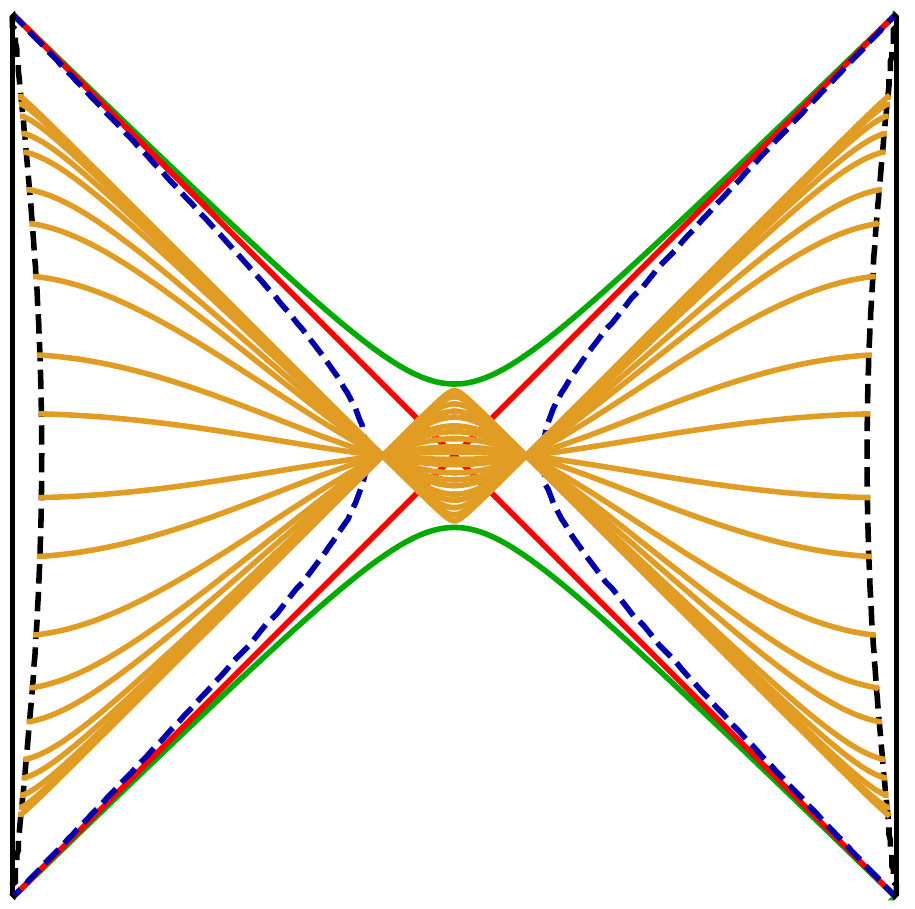} \label{pen_gamma_pos_all}} \quad\quad
                   \subfigure[$\phi_0= -0.85$]{
                \includegraphics[scale=0.4]{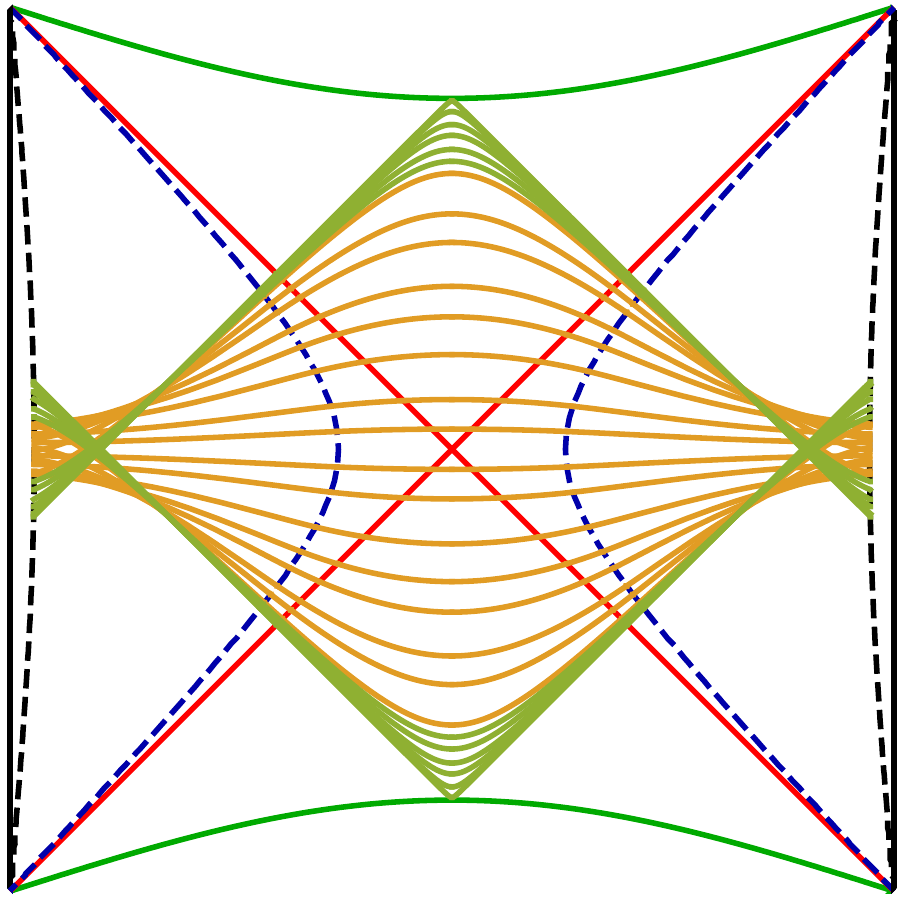} \label{pen_gamma_neg1_all} } \quad\quad
         \subfigure[$\phi_0= -0.95$]{
                \includegraphics[scale=0.4]{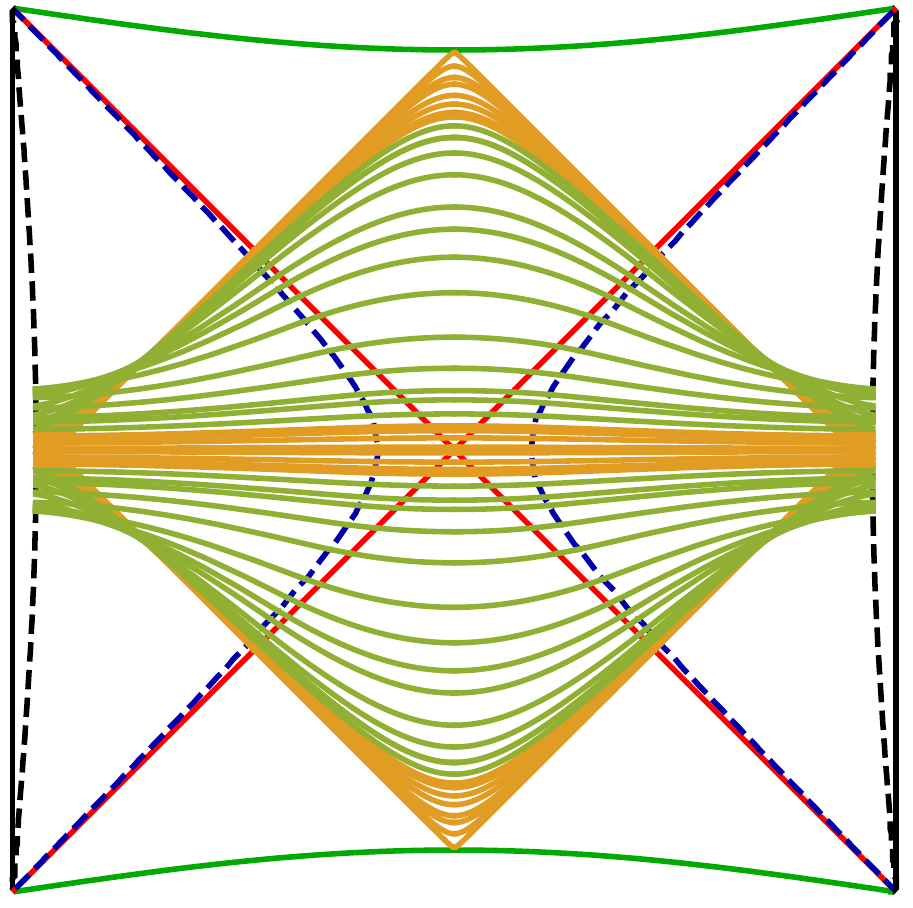} \label{pen_gamma_neg2_all} }
                 \caption{{\footnotesize Same Penrose diagrams as in figure \ref{fig_pen_gamma}, but here we plot geodesics for many different values of P. }}
\label{fig_pen_gamma_all}
\end{figure}

For $-1<\phi_0<-\frac{1}{\sqrt{2}}$, or $-1<\gamma<0$, the situation is more interesting. In that range, at very early times, there are three different maximal geodesics anchored at the very same boundary time. As we will see later, they also have different lengths. Depending on the value of $\phi_0$, there is another range of times where there are two geodesics at the same time --- see figure \ref{t_gamma_neg2}--- or only one --- see figure \ref{t_gamma_neg1}. At larger times, there are no geodesics. Examples of these geodesics in the different Penrose diagrams are shown in figures \ref{fig_pen_gamma} and \ref{fig_pen_gamma_all}.
\newline

\begin{figure}[h!]
        \centering
        \subfigure[]{
                \includegraphics[scale=0.55]{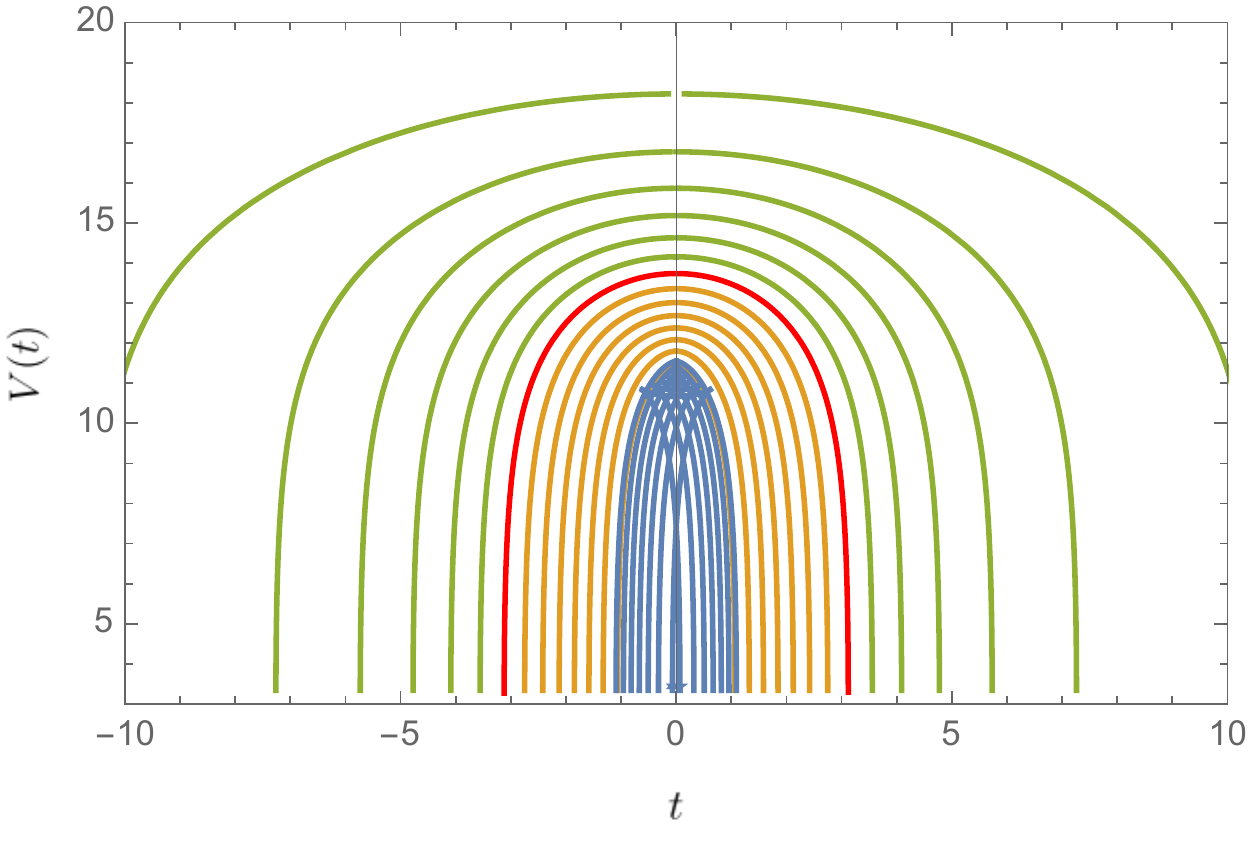} \label{v_t_plot_gamma}} \quad\quad
         \subfigure[]{
                \includegraphics[scale=0.55]{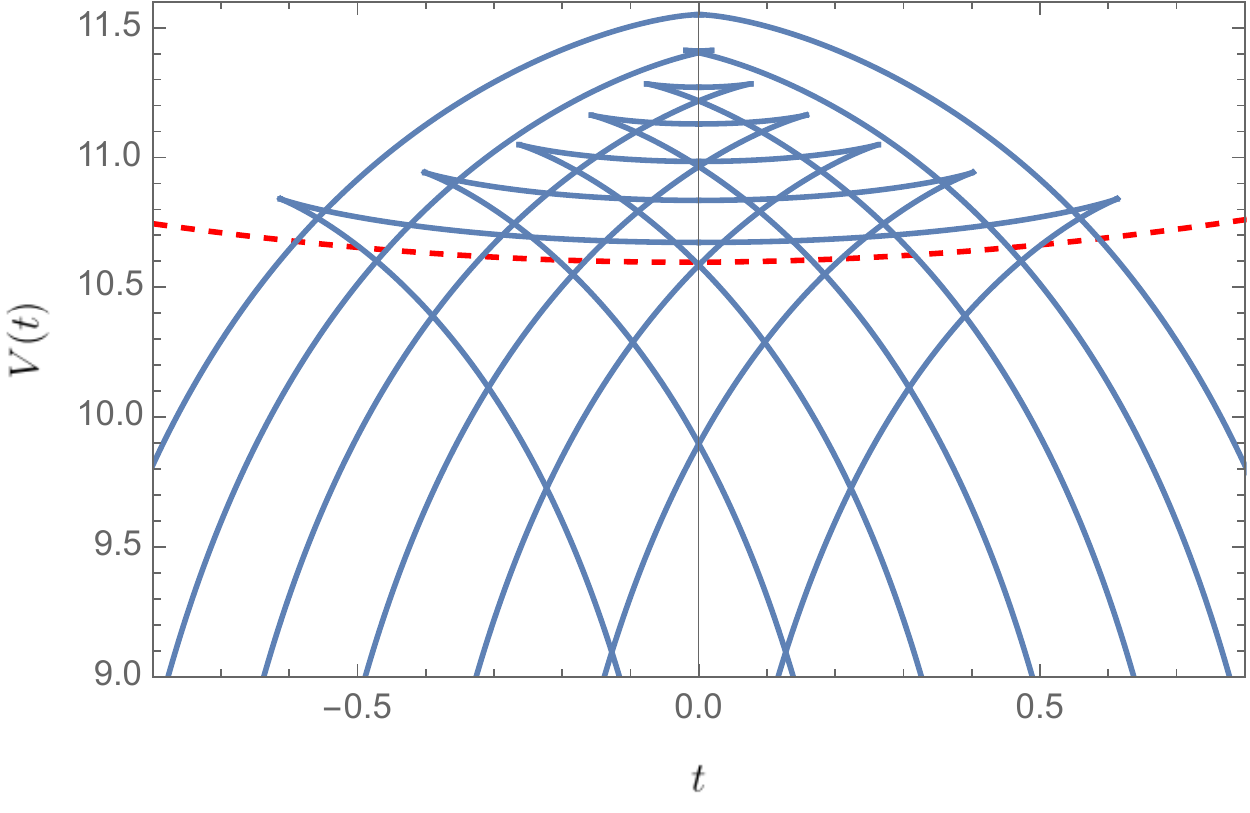}  \label{v_t_plot_gammab2}}
                 \caption{{\footnotesize Volume as a function of the boundary time $t$ for the different $\gamma$-centaur  geometries with $R_b = 100$. In (a), the outermost curve corresponds to $\phi_0=0.7$ and in each curve $\phi_0$ decreases by $-0.1$ until it gets to $\phi_0=-0.7$. The green curves correspond to $\phi_0>0$, and the yellow ones to $\phi_0<0$. In between the curve $\phi_0=0$ is drawn in thicker red. This curve is simply the one found in the previous section. For smaller $\phi_0$ we plotted curves in blue from $\phi_0=-0.74$ to $\phi_0=-0.99$ in steps of $-0.05$. In (b), we zoom in this last region, showing the different behaviour of curves with $\phi_0<-1/\sqrt{2} \sim -0.7$. In dashed red, we show the analytic result for the AdS black hole and we see how as $\phi_0$ approaches $-1$, the curves tend to the one of the black hole, but only for short times, they never reach the linear growth region.
                 }}
\label{fig_gamma_V}
\end{figure}

\noindent {\bf Behaviour of the volume.} In order to obtain the volume as a function of time we need to invert equation (\ref{time_gamma}), obtain $P(t)$, and insert it into equation (\ref{volume_gamma}). However, it  is not possible to do this analytically for arbitrary $\phi_0$. Instead, we will plot the solution parametrically as  $\left( V[P], t [P] \right)$, see  figure \ref{fig_gamma_V}. 

The volume shows two clear different behaviours depending on the value of $\phi_0$. For $|\phi_0|<1/\sqrt{2}$, the form is similar to the one we found for the centaur in the previous section: the volume starts at a maximum at $t=0$ and decreases for some time until there are no more geodesics. For $-1<\phi_0<-1/\sqrt{2}$, the behaviour changes due to the existence of two or three geodesics anchored at the same time. The geodesic with largest volume always shows a characteristic behaviour similar to that of the black hole case but this behaviour does not extend in time until the region where it becomes linear. This is consistent with the fact that the interior geometry does not have a black hole like behaviour and in fact, the geodesics inside the dS region never grow to size larger than $2\pi$. So we do not expect to find linear growth at late times, even though from the outside, most of the geometry looks like the AdS black hole. We return to this point in section \ref{discussion}.

It is possible to find both behaviours analytically close to $t=0$ by expanding the expressions close to $P=0$. The result that we obtain at quadratic order in $t$ is consistent with the expression,
\begin{eqnarray}
V_{\pm}(t) & = & \pi \pm 2 \, \text{arctan} \left( \frac{\sqrt{1-\gamma}}{\sqrt{1+\gamma}} \right) - 2 \log \left( \frac{\sqrt{1+\gamma} \mp \sqrt{1-\gamma}}{\sqrt{2}} \right) + \nonumber \\
& & 2 \log \left( 2 R_b \cos \frac{\sqrt{1 \mp \sqrt{1-\gamma^2}} }{\sqrt{\gamma}} \frac{t}{2} \right) + O(1/R_b) \,, \label{vol_gamma_log_cos}
\end{eqnarray}
where  $V_{\pm}$ corresponds to $\phi_0>0$ and $\phi_0<0$, respectively. 
In the case $-1<\phi_0<-1/\sqrt{2}$ this expression only refers to the uppermost branch of the volume, see figure \ref{v_t_plot_gammab2}. This expression encodes the fact that at early times the volume is quadratic in $t$.
This expression is valid for every $\gamma$ between $-1$ and $1$.
Note that for $\gamma=1$ we recover the centaur result in equation (\ref{vol_t_cent}) and for $\gamma=-1$, the expression becomes that of the AdS black hole -- see equation (\ref{vol_bh_t}). We stress again that though equal, this expression does not hold for arbitrary long times, as in the black hole case. It is interesting to see that for $\gamma<0$, the $\cos$ turns into a $\cosh$, generating the change in behaviour found when going from the yellow curves to the blue ones in figure \ref{v_t_plot_gamma}. Finally, note that as $\gamma \to 0$ with positive $\phi_0\rightarrow{1}/{\sqrt{2}}$, the value of the volume at $t=0$ diverges logarithmically in $\gamma$. This trend is reflected in the uppermost green curves in figure  \ref{fig_gamma_V}. This divergence is independent of the time. Recall that in this case the AdS part of the geometry develops an infinitely long throat, see appendix D of \cite{Anninos:2018svg}.

If more than one geodesic exist at a given time, in order to compute the complexity we need to use the one with maximal volume. If we follow some of the blue curves in figure \ref{v_t_plot_gammab2} along increasing time starting at $t=0$ and always pick the branch of maximal volume, we see two types of behaviours. The maximal volume for values of $\phi_0$ slightly below $-1/\sqrt{2}$ will start increasing following eq. (\ref{vol_gamma_log_cos}). It will then jump discontinuously to a lower value and start deceasing. This decrease will stop at some time of the order of the inverse temperature when the curves stop existing. 

For values of $\phi_0$ closer to $-1$, the maximal volume will be given by eq. (\ref{vol_gamma_log_cos}) (again, just for a short time) and then the curves will stop existing. The transition between the two regimes happens at $\phi_0\sim-0.928$.

\subsection{Geodesics in AdS-to-AdS geometries}
The last case we will analyse is gluing two AdS-like spaces with different radii, in order to highlight the differences with the previous AdS-to-dS case. 

\begin{figure}[h]
	\centering
	\includegraphics[scale=1.2]{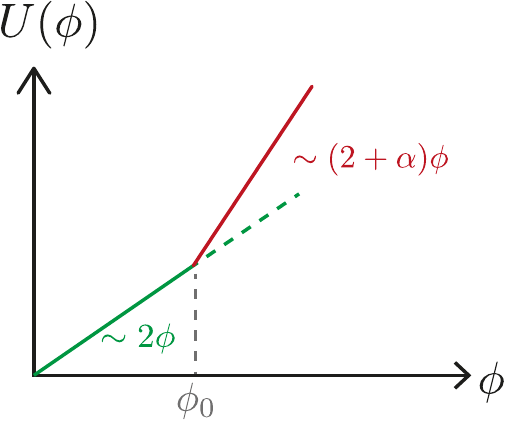}~~~~~~~
	\includegraphics[scale=1.2]{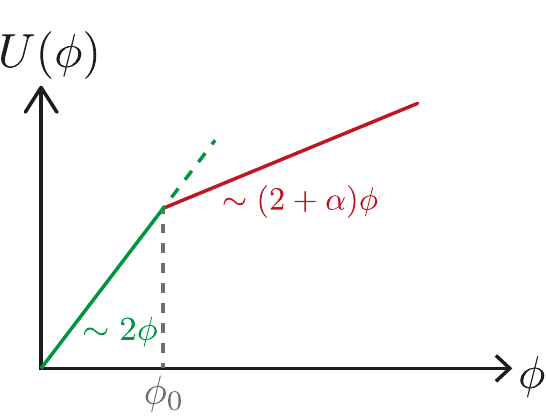}
	\caption{The AdS-to-AdS potential for $\alpha>0$ (left) and $\alpha<0$ (right). The crossing between the two regimes happens at $\phi=\phi_0$.}.
	\label{fig:gcpoten}
\end{figure}

\sloppy
As with the AdS-to-dS case, it is possible to construct dilaton-gravity theories in two dimensions, whose solutions interpolate between two AdS spacetimes with different curvature radii. The dilaton potential is given by
\begin{align}
U(\phi)_{\text{AdS-to-AdS}} &= \begin{cases}
	2\phi \,, & \phi < \phi_0 \,,
\\
	(2+ \alpha) \phi -\alpha  \phi _0 \,, & \phi> \phi_0 \,,
	\end{cases}
\end{align}
in terms of two parameters $\phi_0$ and $\alpha$. The parameter $\phi_0$ fixes the location of the interpolation. We will assume it is greater than the horizon radius ($r_h=1$ in our conventions), so the transition is outside the horizon. The second parameter $\alpha$ characterizes the radius of the second AdS. It ranges between $-2<\alpha<\infty$, where $\alpha = -2$ corresponds to an interpolation to flat spacetime and $\alpha = 0$ corresponds to no interpolation at all, see figure \ref{fig:gcpoten}.

The interior AdS has unit radius and we will set the horizon at $r_h=1$. Then the metric becomes,
\begin{align}
f(r)_{\text{AdS-to-AdS}} &= \begin{cases}
	(r^2 - 1) \,, & 0 < r < \phi_0 \,,
\\
	r^2 -1 +\frac{1}{2} \alpha  (r- \phi_0)^2 \,, & \phi_0 < r < \infty \,.
	\end{cases}
\end{align}

\begin{figure}[h]
	\centering
	\includegraphics[scale=0.7]{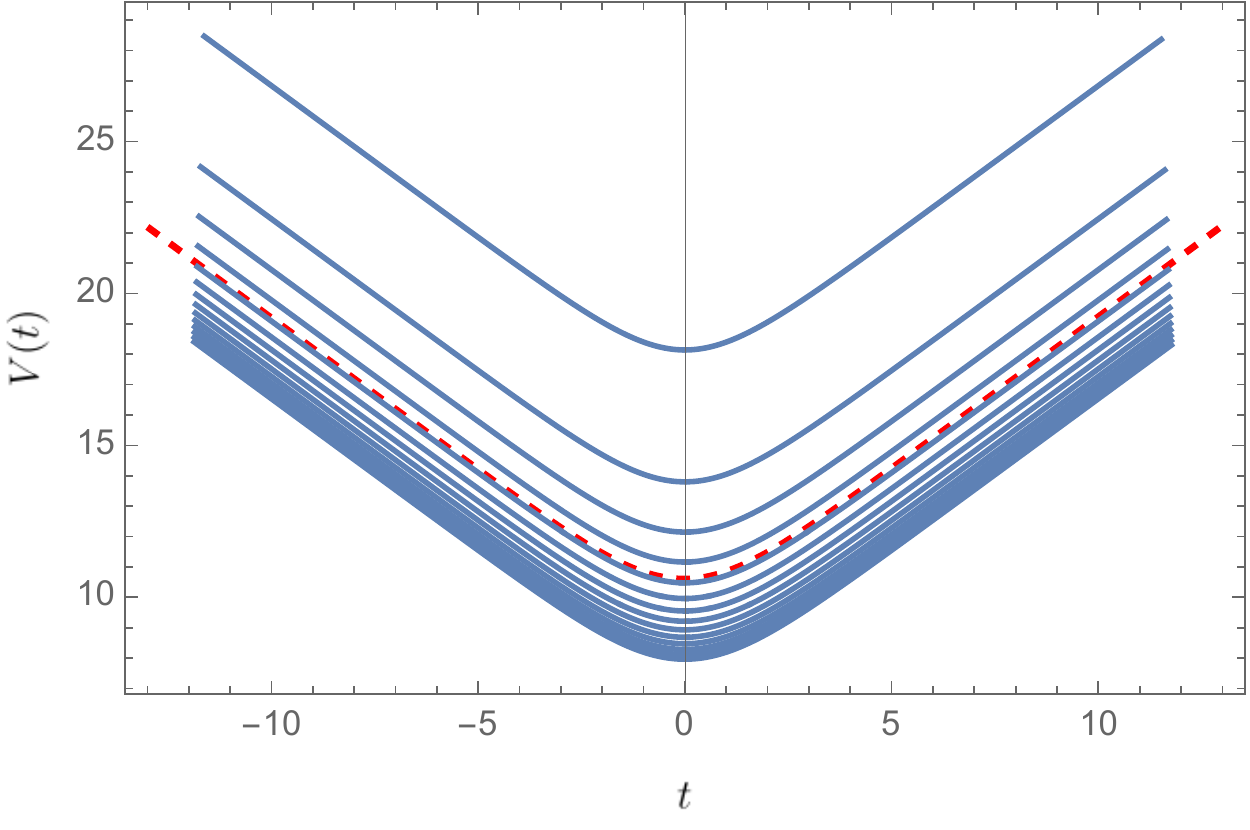}
	\caption{Volume as a function of boundary time $t$ for the different AdS-to-AdS geometries with $R_b = 100$ and $\phi_0=2$. The curves go from $\alpha = -1.9$ at the top to $\alpha=5$ at the bottom, in steps of 0.5. In all cases, for long times, the behaviour is linear with the same slope. In dashed red, we show the curve for the AdS black hole,  corresponding to the case $\alpha=0$. }
	\label{fig:ads_ads}
\end{figure}

The procedure is identical to the other cases, so we will just state here the main results. The turning point is
\begin{equation}
r_t = \sqrt{1-P^2} \,,
\end{equation}
corresponding to a range $-1<P<1$ of the conserved momentum. The tortoise coordinate is
\begin{align}
r^*(r)_{\text{AdS-to-AdS}} &= \begin{cases}
	-\frac{1}{2} \log \left| \frac{ r+1 }{r-1 }\right| + c_1  \,, & 0 < r < \phi_0 \,,
\\
	\frac{\text{arctan} \left(\frac{(\alpha +2) r-\alpha  \phi _0}{\sqrt{2 \alpha  \left(\phi _0^2-1\right)-4}}\right)-\pi/2}{\sqrt{\frac{1}{2} \alpha  \left(\phi _0^2-1\right)-1}}  \,, & \phi_0 < r < \infty \,,
	\end{cases}
\end{align}
where the constant $c_1$ is given by
\begin{equation}
c_1 =  \frac{2 \,  \text{arctan} \left(\frac{\phi _0}{\sqrt{\frac{1}{2} \alpha  \left(\phi _0^2-1\right)-1}}\right) - \pi}{\sqrt{2(\alpha  \phi _0^2-\alpha -2)}}- \frac{1}{2} \log \left(\frac{\phi _0-1}{\phi _0+1}\right)  \,.
\end{equation}
At the turning point, this becomes
\begin{equation}
r_t^* =  \frac{ \,  \text{arctan} \left(\frac{\phi _0}{\sqrt{\frac{1}{2} \alpha  \left(\phi _0^2-1\right)-1}}\right) - \pi/2}{\sqrt{\frac{1}{2} \alpha  \left(\phi_0^2-1\right)-1}} + \text{arctanh} \left(\frac{\sqrt{1-P^2} \phi _0-1}{\sqrt{1-P^2}-\phi _0}\right) \,.
\end{equation}
The volume integral gives
\begin{equation}
V[P] = \log \left(\frac{\phi_0 + \sqrt{P^2+\phi _0^2-1}}{\phi _0-\sqrt{P^2+\phi _0^2-1}}\right) + \frac{2 \sqrt{2} \log \left(\frac{-\alpha  \phi _0+\sqrt{\alpha +2} \sqrt{2 \left(P^2+R_b^2-1\right)+\alpha  \left(R_b-\phi _0\right){}^2}+(\alpha +2) R_b}{\sqrt{2} \sqrt{(\alpha +2) \left(P^2+\phi _0^2-1\right)}+2 \phi _0}\right)}{\sqrt{\alpha +2}} \,.
\end{equation}
It is straightforward to obtain an expression for the times for any value of $\alpha$, $\phi_0$ and for large $R_b$
\begin{equation}
\begin{split}
t=
&
\log \left(\frac{\left(\phi _0-1\right) \left(\mu +P \phi _0\right)}{\left(\phi _0+1\right) \left(\mu -P \phi _0\right)}\right)+2\text{arccoth}\left(\phi _0\right) 
\\
&
+2\sqrt{\frac{2}{\nu }} \,\text{arccoth} \left(\frac{\sqrt{\alpha +2} P}{\sqrt{\nu }}\right)- \sqrt{\frac{2}{\nu }}\text{arccoth}\left(\frac{\mu(\alpha +2)   P}{ \sqrt{2\nu } \,\phi _0-\nu +(\alpha +2) P^2}\right)
\\
&
-\sqrt{\frac{2}{\nu }}\text{arccoth}\left(\frac{\mu (\alpha +2)  P}{ \sqrt{2\nu } \,\phi _0+\nu - (\alpha +2) P^2}\right)
+ O(1/R_b^2),
\end{split}
\end{equation}
where we have defined $\nu \equiv \alpha(1-  \phi_0^2)+2$, $\mu \equiv \sqrt{P^2 + \phi_0^2-1}$.
It is possible to get analytically the form of $V(t)$, expanding both $V[P]$ and $t[P]$ close to $P=\pm1$. This yields a surprisingly simple expression
\begin{equation}
V(t) = \frac{2 \sqrt{2} \log  R_b}{\sqrt{\alpha +2}} +  t + ... \,,
\end{equation}
where independently of $\alpha$ and $\phi_0$, the volume grows linearly in time with coefficient 1. This can be further seen in figure \ref{fig:ads_ads}, where we plot the volume for different values of $\alpha$, keeping $\phi_0$ fixed. We see that in all cases, the volume grows linearly in time, as in the AdS black hole, which is  strikingly different from the cases with dS interiors.

\section{Discussion} \label{discussion}

In this paper, we computed the length of spacelike geodesics in different spacetime geometries in two dimensions. In most cases, the geometries analysed are two-sided, asymptotically AdS geometries with horizons in the interior. The nature of the horizon can be rather different, with the spacetime contracting (as in a black hole) or expanding (as in cosmology) towards future/past infinity. The results are significantly different depending on each situation. Mainly, while in the black hole case the length grows linearly in time for long times, in most cases of the cosmological scenario, the geodesics decrease in length for a short time of the order of the inverse temperature and then they stop existing. These interesting feature of cosmological geodesics raises a number of  interesting possibilities to be discussed.
\newline\newline
{\textbf{The complexity equals volume conjecture.}} In the black hole case, spacelike geodesics anchored at the same time on both boundaries always exist. Therefore, the prescription to define complexity as the volume of these geodesics seems reasonable. In some of the geometries analysed in the present paper this is not the case: geodesics anchored at the same time only exist for a short time of the order of the inverse temperature. However, from a boundary perspective, we do not expect that after some time complexity stops being defined. On the contrary, it has been proposed as a measure for late-time entanglement properties. One possibility to solve this apparent contradiction is to consider geodesics that  go through future/past  infinity which will have infinite length \cite{Susskind:2021esx}. In this case, we expect the validity of the semi-classical  approximation to be lost, but the behaviour of the length is nevertheless intriguing. It decreases for a short time and then jumps to infinity instantaneously. This resonates with the idea of ``hyperfast'' complexity growth recently explored in \cite{Susskind:2021esx}.  
In that paper, the geodesics in pure dS were considered  where a temporal boundary is absent. Therefore, the complexity was associated with a notion of time evolution directly on the dS horizon. The interpretation suggested for the complexity becoming infinite after times of the order of the inverse temperature was that of a model whose Hamiltonian couples a significant portion of the system's degrees of freedom within each of its terms. Our approach utilizes a different notion of time evolution and it will be interesting to understand the relation between the two. It is curious to note that if the dual to boundary complexity is instead assumed to be the volume that lies only behind the horizon as in \cite{Susskind:2021esx}, we find linear growth behind the dS horizon as shown in equation  \eqref{volume_inside_centaur}. This is valid only for times of the order of the inverse temperature whereas the black hole volume grows quadratically with time at such early times, as can be checked from expanding equation \eqref{vol_bh_t}.


A different possibility is that \emph{complexity=volume is not enough} after all, and we need another prescription to compute the complexity of the boundary state. It would be interesting to compare the results obtained for the volume with the ones given by other proposals such as the complexity=action -- see \cite{Brown:2018bms, Goto:2018iay} for computation in JT gravity --, and complexity=spacetime-volume \cite{Couch:2016exn}. Those two last proposals are presumably well defined for all boundary times (but possibly also yield divergent answers).

Even though the calculations in the present paper only involve geometries with trivial topology, it is also interesting to compare our results with the recent non-perturbative definition of length proposed in \cite{Iliesiu:2021ari}. Under certain assumptions for the potential $U(\phi)$ (see \cite{Maxfield:2020ale, Witten:2020wvy}), the authors of \cite{Iliesiu:2021ari} find a universal linear growth in the length for times between the thermalization time and $e^{S_0}$, independent of the form of the potential. It would be interesting to understand how the arguments of \cite{Iliesiu:2021ari} break down in our case.
\newline\newline
{\textbf{Shockwaves and out-of-time-ordered correlators.}} Another interesting observable that probes quantum chaos is the out-of-time-ordered correlator (OTOC). It has been shown that both the exponential growth of the OTOC and the linear growth of complexity are related to the chaotic nature of the system under consideration \cite{Shenker:2013pqa,Stanford:2014jda}. As mentioned, the OTOC in an interpolating geometry with a dS horizon does not exhibit exponential growth \cite{Anninos:2018svg}. Another evidence for the unusual behaviour of the system is found in the results of the present paper where the length does not grow linearly with time for long times. Close to the boundary, these geometries have a Schwarzian-like behaviour, governed by the following action \cite{Anninos:2018svg, Jensen:2019cmr, Yoon:2019cql, Nayak:2019evx}, 
\begin{equation}
S_{\text{bdy}} = \frac{\phi_b}{8 \pi G_N} \int du  \left( \frac{\gamma}{2} (\partial_u \tau(u))^2  - \text{Sch} [\tau(u),u] \right) \,,
\end{equation}  
which might suggest that the OTOC behaves like $\sim \cos \sqrt{\gamma} t$, giving exponential growth for negative $\gamma$ geometries. However, the results obtained in this paper show that there is no linear growth for the geodesic length even in that case. This is due to the fact that the horizon interior is filled with dS spacetime. If a precise relation between complexity and the OTOC is established this would suggest that interesting cancellations should   happen in the OTOC between the boundary Schwarzian action and the interior modes. It would be interesting to confirm this fact by either doing a direct calculation of the OTOC following \cite{Anninos:2018svg} or by computing the geodesic length in a shockwave setup \cite{Stanford:2014jda,Chapman:2018dem, Chapman:2018lsv}.

The volume in shockwave geometries was related to the complexity growth of the precursor operator encoding the influence of a perturbation inserted some time $t_w$ in the past on the system. In terms of $t_w$, the complexity grows initially exponentially according to the Lyapunov exponent  of maximally chaotic systems $\lambda_L = 2\pi T$ \cite{Maldacena:2015waa} and then at the scrambling time  $t_* = (1/2\pi T)\log S$ \cite{Shenker:2013pqa}, it starts growing linearly at a rate which is twice the usual linear growth of complexity in the black hole background. It would be useful to study this observable in the centaur geometries to characterize chaos and scrambling in those systems. 
\newline\newline
{\textbf{Complexity of formation.}}
The length of our geodesic is regulated by the finite location of the boundary. This has a large effect on the complexity, but this large effect does not depend on time. It was suggested in 
\cite{Chapman:2016hwi} that a useful way to subtract the effect of the boundary, which functions here as a UV regulator, is to consider differences between the complexity of our system and a reference system which was taken to be the AdS vacuum. This vacuum-subtracted complexity goes under the name of complexity of formation
\begin{equation}
\Delta\mathcal C = \mathcal{C} -  \mathcal{C}_{\text{global}} =  \frac{\Phi_0}{G_N \ell} \left(V -  V_{\text{global}}\right),
\end{equation}
where as opposed to the higher dimensional case \cite{Chapman:2016hwi}, here we subtract a single copy of AdS$_2$ since global  AdS$_2$ geometry has two boundaries.
Evaluating this for the AdS$_2$ black hole, we find using equation \eqref{vol_bh_t}
\begin{equation}
\Delta\mathcal C_{BH}(t) = 8 S_0 \log \left( \cosh (\pi Tt) \right) + O(1/R_b) \,,
\end{equation}
which simply vanishes for $t=0$ because it is essentially the same space. For the centaur, we find using  equation \eqref{vol_t_cent}
\begin{equation}\label{forNetta}
\Delta\mathcal C_{\text{centaur}}(t) =\pi+ 8 S_0 \log \left( \cos (\pi Tt) \right) + O(1/R_b), \qquad |t|T<1/2 \,, 
\end{equation}
at $t=0$ it gives $\pi$ which is exactly the length in the dS$_2$ part as we could have expected. Similarly to the higher dimensional case, also here the difference in complexities (at $t\neq 0$) grows linearly with the thermal entropy.  
It is curious to note that the two formulas are related by analytic continuation, but only for short times.
Although the volume is always positive, note that the expression \eqref{forNetta} can become negative. This contrasts with the theorem proven in \cite{Engelhardt:2021mju} for asymptotically AdS spacetimes in four dimensions. It would be interesting to use our example to find limitations on possible generalizations of this theorem to different dimensions.
\newline\newline
{\textbf{Towards a microscopic quantum theory for dS.}} If the volume is connected to some notion of complexity, then it would be interesting to see what is needed from the boundary perspective to obtain this type of complexity. We have already a lot of evidence showing that the cosmological horizon is not a quantum maximally chaotic horizon like the black hole one. Given we are studying the problem in two dimensions, the dual theory will be quantum mechanics and we might be able to compute complicated quantities such as the complexity or the OTOC in microscopic models or even try to reverse-engineer models to behave in accordance with our holographic results, possibly along the line of \cite{Anous:2021eqj} or the SYK RG-flows of \citep{Anninos:2020cwo}.
\newline\newline
{\textbf{Complex geodesics and dS 2-pt function.}} Finally, it is interesting to discuss the geodesics in pure dS. In particular, the length of the geodesics should be related to the two-point function of heavy fields in dS through a saddle-point approximation. The two-point function is known for scalar fields of arbitrary mass at any two points \cite{Spradlin:2001pw}. However, we showed that there are no real, finite, geodesics anchored at any time different than $t=0$. It is possible that the saddle-point geodesic becomes complex or that from some reason the saddle-point approximation fails in this case. In any case, we hope to address this question in future work.

\section*{Acknowledgements}

We gratefully acknowledge discussions with D. Anninos and L. Iliesiu. The work of SC is supported by the Israel Science Foundation (grant No. 1417/21). SC acknowledges the support of Carole and Marcus Weinstein through the BGU Presidential Faculty Recruitment Fund.
The work of DAG is funded by the Royal Society under the grant ``The Resonances of a de Sitter Universe'' and the ERC Consolidator Grant N. 681908, ``Quantum black holes: A microscopic window into the microstructure of gravity''. EDK is supported by the Israel Science Foundation (grant No. 1111/17) awarded to Eric Kuflik.

\appendix

\section{Thermodynamics of dilaton-gravity theories and $\gamma$-centaur geometries} \label{app_thermo}

In this appendix, we study the thermodynamics of dilaton-gravity theories with a general dilaton potential. This has been studied previously in a variety of different contexts, including \cite{Cavaglia:1998xj, Grumiller:2007ju, Anninos:2017hhn, Witten:2020ert}. For any continuous dilaton potential, the temperature, entropy and specific heat are given by
\begin{equation}\label{thermo}
T = \frac{U(\phi(r_h))}{4\pi}~,  \quad\quad  S = \frac{\Phi_0 + \phi(r_h)}{4 G_N} ~, \quad\quad C = \frac{1}{4 G_N} \frac{U(\phi(r_h))}{\partial_\phi U(\phi(r_h))}~,
\end{equation}
where we have set $\ell=1$.\footnote{In this appendix we  introduce back the factors of $r_h$. This is important for computing the thermodynamic quantities.} The thermal partition function and energy are given by 
\begin{equation}
\log Z = S - \frac{E}{T}~, \quad\quad E = -\frac{1}{16 \pi G_N}\int_{r_h}^{R_b} d\tilde{r} \, U(\phi(\tilde{r}))~,
\end{equation} 
where $R_b$ is the location of the Euclidean AdS$_2$ boundary. The energy diverges in the limit $R_b \to \infty$. This divergence can be absorbed in an infinite shift of the ground state energy. It is more meaningful to consider the energy difference between two solutions
\begin{equation}
\Delta E = E_{r_h} - E_{r_h'} = \frac{1}{16 \pi G_N}\int_{r_h'}^{r_h} d\tilde{r} \, U(\phi(\tilde{r}))~.
\end{equation}

\begin{figure}[h]
	\centering
	\includegraphics[scale=1]{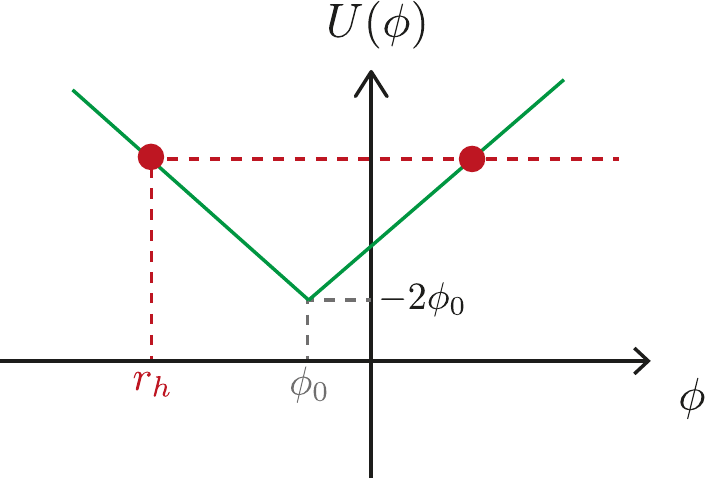}~~~~~~~
	\includegraphics[scale=1]{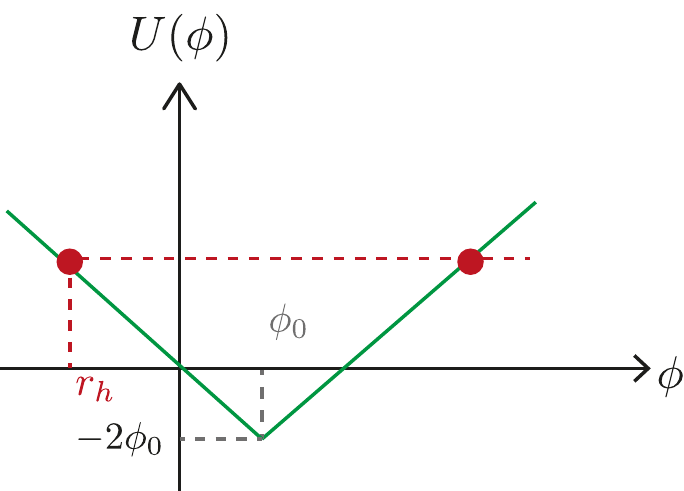}
	\caption{The $\gamma$-centaur potential for $\phi_0<0$ (left) and $\phi_0>0$ (right). The region where the slope is negative describes dS while the region where the slope is positive describes AdS and the crossing between the two  regions happens at $\phi=\phi_0$. There are at most two different values of $r_h$ corresponding to a given temperature $T=U(\phi(r_h))/4\pi$ indicated by red dots in this illustration. To have interpolating solutions we need $r_h<\phi_0$.}
	\label{fig:gc2poten}
\end{figure}

We obtain the thermodynamics for the interpolating geometries studied in the main text by choosing the appropriate dilaton potentials. Let us consider $U(\phi) = 2 (|\phi - \phi_0| - \phi_0)$, see figure \ref{fig:gc2poten}. 

For fixed $\phi_0$, there are at most two values of $r_h$ corresponding to any given temperature. The metric \eqref{blackfactor} obtained with the choice $r_h>\phi_0$, will have $R=-2$ everywhere. If in addition we have $\phi_0>0$, we have solutions for all $T \geq 0$. On the other hand, for $\phi_0<0$, our solutions will always have $T\geq T_{\text{min}} = |\phi_0|/2\pi$, so there are no zero-temperature solutions in this case. The specific heat of all solutions with $r_h>\phi_0$ is given by $C(T) = \pi T/2 G_N$.

To have an interpolating solution such as the ones analysed in the main text, we need $r_h<\phi_0$. Using eq.~\eqref{blackfactor}, we can show that in this case, the metric is given by\footnote{Note, that the $\phi_0$ used  here is $|r_h|$ times the $\phi_0$ used in the main text. We keep using $\ell=1$.}
\begin{align}
f(r)_{\gamma} &= \begin{cases}
	r_h^2- r^2 \,, & -\infty < r < \phi_0 \,,
\\
	r_h^2+ r^2+2 \phi _0 \left(\phi _0-2 r\right)  \,, & \phi_0 < r < \infty \,.
	\end{cases}
\end{align}
When $\phi_0>0$, we have to make sure that $r_h$ is chosen such that the metric is everywhere positive, see right panel of figure \ref{fig:gc2poten}. This restricts $r_h<-\sqrt{2}\,\phi_0$. 
For $\phi_0<0$, the metric will be everywhere positive, but we still have to make sure that $r_h<\phi_0$. Combining these two conditions, we find $r_h<\phi_0<\frac{1}{\sqrt{2}}|r_h|$ which upon setting $r_h=-1$ restores the range of  $\phi_0$ in the main text.  
In all interpolating solutions, the temperature is given by
\begin{equation}
T = \frac{|r_h - \phi_0| - \phi_0}{2 \pi} = \frac{|r_h|}{2\pi} \,.
\end{equation}
Note that from the above constraints it follows that for $\phi_0<0$, there only exist interpolating solutions with $T> |\phi_0|/2\pi$ while for $\phi_0>0$ solutions exist only for $T>\phi_0/\sqrt{2}\pi$.  

It is straightforward to compute the different thermodynamic quantities for the interpolating geometries. The entropy and the specific heat are given by 
\begin{equation}
S(T)  = \frac{\Phi_0 - 2 \pi T}{4 G_N} \,\,\,\,\, \,, \,\,\,\,\, C(T) = -\frac{\pi  T}{2 G_N} \,,
\end{equation}
where it should be noted that the minus sign comes from having $r_h<0$. There are also solutions with positive specific heat, but these have a dilaton that decreases towards the boundary, see \cite{Anninos:2017hhn, Anninos:2018svg}. An important feature of these geometries is that the specific heat is linear in the temperature. The energy is given by,
\begin{equation}
E(T) = -\frac{1}{16 \pi  G_N} \left( R_b^2-4 \phi _0 R_b+2 \phi _0^2+4 \pi ^2 T^2 \right) \,,
\end{equation}
which as mentioned earlier diverges as $R_b \to \infty$. Nevertheless, only the last term depends on the temperature. Note, that this term is finite and does not depend on $\phi_0$. The regulated energy is $\Delta E = -\frac{\pi}{4G_N} \Delta (T^2)$. Note that apart from the ground state energy, thermodynamic quantities do not depend on $\phi_0$.

\section{From conformal coordinates to the Schwarzschild gauge} \label{app_coordinates}

In the main text we use mostly Schwarzschild coordinates for the metric. This is a useful gauge to compute the length of geodesics. For the $\gamma$-centaur geometries, though, it is sometimes useful to go to the conformal gauge. In this appendix, we show how to go between the two coordinate systems. We start with the metric that was used in the AdS portion of the $\gamma$-centaur geometries in section \ref{gamma_cen_sec}: 
\begin{equation}
ds^2_{\gamma} = - f(r)_\gamma dt^2 + \frac{dr^2}{f(r)_\gamma} \,\,\,\,\, \,, \,\,\,\,\, f(r)_\gamma = 1+ r^2 +2 \phi_0( \phi_0- 2r) \,.
\end{equation}
Note that this can be written as
\begin{equation}
f(r)_\gamma = (r- r_+)(r+r_-) \,\,\,\,\, \text{with} \,\,\,\,\, r_\pm = 2 \phi_0 \pm \sqrt{-1+2\phi_0^2} \,.
\end{equation}

We can define $\tilde{r} = r - 2 \phi_0$. In terms of this coordinate, $f(\tilde{r}) = \tilde{r}^2 + (1-2 \phi_0^2)$. We want to compare this metric to the one in conformal coordinates. The general negative curvature solution is given by
\begin{equation}\label{appB}
ds^2 = \frac{\gamma}{\sin^2 \sqrt{\gamma} \rho} (-dt^2 + d\rho^2) \,.
\end{equation}
We can always rescale $\rho$ and $t$ and  \eqref{appB} will locally look like the AdS$_2$ black hole metric. However, to fix the temperature of the flow geometries, we fix the periodicity of the time coordinate (in Euclidean signature) using the dS metric, so, globally, all these $\gamma$-centaur geometries are different \cite{Anninos:2018svg}. 

It is now straightforward to perform the change of coordinates from Schwarzschild to conformal gauge by identifying
\begin{equation}
\tilde{r} = \sqrt{\gamma \left(\csc ^2\left(\sqrt{\gamma } \rho \right)-1 \right)} \\,\,\,\, \text{with} \,\,\,\,\, \gamma \equiv 1- 2\phi_0^2 \,,
\end{equation}
where you might identify the last parameter as it has been extensively used across section \ref{gamma_cen_sec}.

\bibliographystyle{JHEP}

\bibliography{bibliography}

\end{document}